# Spatiotemporal torquing of light

S.W. Hancock, S. Zahedpour, A. Goffin, and H.M. Milchberg

*Institute for Research in Electronics and Applied Physics, University of Maryland, College Park, Maryland 20742, USA*

We demonstrate the controlled spatiotemporal transfer of transverse orbital angular momentum (OAM) to electromagnetic waves: the *spatiotemporal* torquing of light. This is a radically different situation than OAM transfer to longitudinal, spatially-defined OAM light by stationary or slowly varying refractive index structures such as phase plates or air turbulence. We show that transverse OAM can be imparted to a short light pulse only for (1) sufficiently fast transient phase perturbations overlapped with the pulse in spacetime, or (2) energy removal from a pulse that already has transverse OAM. Our OAM theory for spatiotemporal optical vortex (STOV) pulses [Phys. Rev. Lett. **127**, 193901 (2021)] correctly quantifies the light-matter interaction of this experiment, and provides a torque-based explanation for the first measurement of STOVs [Phys. Rev. X **6**, 031037 (2016)].

## I. INTRODUCTION

The study and applications of light carrying longitudinal orbital angular momentum (OAM) have been actively pursued since it was realized that Laguerre-Gaussian ($LG_{pm}$) modes with integer radial and azimuthal indices $p$ and $m$ have an OAM of $m\hbar$ per photon directed parallel or anti-parallel to the propagation axis [1]. Whether or not the OAM content was directly important to applications, OAM-carrying light such as $LG_{pl}$ and Bessel-Gaussian ($BG_m$) modes with $m \neq 0$ has found uses in areas such as optical trapping [2], super-resolution microscopy [3], generation of long air waveguides [4], and plasma waveguides [5]. Other proposed uses of longitudinal OAM beams include free-space communications [6-8], quantum key distribution [9], and generating large magnetic fields in intense laser-plasma interaction [10].

That light could carry OAM oriented transverse to its propagation direction was first revealed in a high field nonlinear optics experiment [11]. The transverse OAM density was carried by spatiotemporal optical vortices (STOVs)—vortices embedded in spacetime— generated by the extreme spatiotemporal phase shear produced in the filamentation and self-guiding of intense femtosecond laser pulses in air [11]. STOVs are naturally emergent and necessary electromagnetic structures that govern optical energy density flow during self-guided propagation, and are a universal consequence of any arrested self-focusing process such as relativistic self-guiding in plasmas [12]. As they are carried by short pulses and are of finite duration, these structures are necessarily polychromatic [13]. The realization that STOVs were generated by phase shear in spacetime [11] led to a method to generate them linearly and controllably, using a $4f$ pulse shaper to apply shear in the spatiospectral domain and then return the pulse to the spatiotemporal domain [14-16]. A new single shot diagnostic, TG-SSSI (transient grating single shot supercontinuum spectral interferometry) [17] captured the free-space propagation of pulse-shaper-generated STOV pulses from the near to far field. Later work used a pulse shaper to generate STOVs measured in the far field only [18]. Since then, alternative methods for STOV generation have been proposed [19-22], and calculation of higher order STOV propagation has been performed [23]. In further work, it was verified that transverse OAM is carried by photons, in experiments demonstrating OAM conservation under second harmonic generation (SHG) [24-26]. Simulations have also predicted the generation of high harmonic STOV photons [27].

Despite the rapidly increasing experimental activity studying STOVs, there had been no theoretical analysis of their OAM content until recently [28-30], where one result determined that STOV-based OAM must take half-integer values [30], with the other claiming that only integer values are allowed [28,29] (see further discussion in Appendix A). This difference is more than just an academic question, as it quantifies the exchange of transverse OAM in light-matter interactions. Such interactions, the subject of this paper, are one of the building blocks of future applications of STOVs.

Interactions of longitudinal OAM-carrying beams with matter have been long studied. One early example is the interaction of a $LG_{0m}$ donut mode with a macroscopic particle, causing it to rotate about the OAM axis [31]. The converse of this process can be viewed as the torquing of light, in which a light beam gains OAM from an interaction with matter. A simple example of this is the pickup of OAM by a beam passing through a spiral phase plate [5] or any refractive index structure that imparts a non-zero azimuthal phase shift about the propagation axis. For example, $LG_{0m}$ donut beam propagation through a turbulent atmosphere leads to an output beam carrying a spectrum of longitudinal OAM states $m, m \pm 1, m \pm 2, ....$ [6, 32], owing to the random azimuthal phase shifts picked up over the propagation range. In all of these cases, the OAM beam can be CW and monochromatic, with the refractive index structures static on the timescale of the beam evolution: such torquing of light makes preservation of pure longitudinal OAM states difficult.

In this paper, we present the first experimental evidence of the controlled spatiotemporal transfer of *transverse* OAM to light by matter: the *spatiotemporal* torquing of light. This is a radically different situation than the torquing of longitudinal, spatially-defined OAM light by stationary or slowly varying refractive index structures such as phase plates or air turbulence. We demonstrate that transverse OAM of a light pulse can be changed only for sufficiently fast transient phase perturbations that overlap with the pulse in spacetime, or by removing energy from a pulse already possessing transverse OAM. We explore the physics of what constitutes an optimal overlap. Furthermore, we experimentally verify our "half-integer" theory of STOV OAM [30]; the theory is crucial to correctly quantifying the light-matter interaction of this experiment. We also make a connection to the first measurement of STOVs [11], providing a spatiotemporal torque-based explanation for their generation.

## II. DETERMINING CHANGES IN TRANSVERSE ORBITAL ANGULAR MOMENTUM

The perturbation-induced change in the orbital angular momentum of an optical pulse can be determined from measurements of the amplitude and phase of the pulse before and after the perturbation. For the well-known case of pulses with longitudinal OAM, say along $\hat{\mathbf{z}}$, the procedure is straightforward: If the pre- and post-perturbation complex electromagnetic field envelopes are $E_s$ and $E_{sp}$, then the change in longitudinal OAM per photon (which is necessarily *intrinsic* OAM) is computed as $\Delta\langle L_z\rangle = \langle L_z\rangle_{sp} - \langle L_z\rangle_s = u_{sp}^{-1}\langle E_{sp}|L_z|E_{sp}\rangle - u_s^{-1}\langle E_s|L_z|E_s\rangle$, where $L_z = (\mathbf{r}\times\hat{\mathbf{p}})_z = -i(x\,\partial/\partial y - y\,\partial/\partial x)$ is the longitudinal OAM operator. Here we use the linear momentum operator $\hat{\mathbf{p}} = -i\nabla$, and $u_{s,sp} = \langle E_{s,sp}|E_{s,sp}\rangle = \int d^3\mathbf{r}\,|E_{s,sp}|^2$. The expectation values of $L_z$ for the pre- and post-perturbation fields are $\langle L_z\rangle_{s,sp} = u_{s,sp}^{-1}\langle E_{s,sp}|L_z|E_{s,sp}\rangle = u_{s,sp}^{-1}\int d^3\mathbf{r}\,E_{s,sp}^* L_z E_{s,sp}$, where the integrals are taken over all space, with $d^3\mathbf{r} = dxdydz$ and the origin taken as the energy density centroid (or "centre of energy"). This choice of origin isolates the intrinsic OAM from extrinsic OAM; it is further discussed in Appendix A. The same result is obtained by directly integrating the OAM density of the fields [33]: $\Delta\langle L_z\rangle = 2k_0 U_{sp}^{-1}\int d^3\mathbf{r}\,[(\mathbf{r}-\mathbf{r}_{sp})\times(\mathbf{E}_{sp}\times\mathbf{H}_{sp}^*)]_z - 2k_0 U_s^{-1}\int d^3\mathbf{r}\,[(\mathbf{r}-\mathbf{r}_s)\times(\mathbf{E}_s\times\mathbf{H}_s^*)]_z$, where $U_{s,sp} = \int d^3\mathbf{r}\,(|\mathbf{E}_{s,sp}|^2 + |\mathbf{H}_{s,sp}|^2)$, $\mathbf{r}_{s,sp} = U_{s,sp}^{-1}\int d^3\mathbf{r}\,\mathbf{r}(|\mathbf{E}_{s,sp}|^2 + |\mathbf{H}_{s,sp}|^2)$ are the respective pulse centres of energy, $\mathbf{H}_{s,sp}$ is the magnetic field, and $k_0$ is the wavenumber of the fields, which can be monochromatic. Here we have assumed propagation in a dilute, non-magnetic material with index of refraction satisfying $Re(n) \cong 1$.

Likewise, for changes in transverse spatiotemporal OAM, an operator-based calculation should agree with a direct field-based calculation using the transverse OAM density. That is, if $E_s$ and $E_{sp}$ are pre- and post-perturbation $\hat{\mathbf{y}}$-polarized pulses propagating along $\hat{\mathbf{z}}$ with transverse OAM oriented along $\hat{\mathbf{y}}$ (ensuring no effects of spin angular momentum), the change in intrinsic transverse OAM per photon, $\Delta\langle L_y\rangle$, should be calculable either as

$$\Delta\langle L_y\rangle = \langle L_y\rangle_{sp} - \langle L_y\rangle_s = u_{sp}^{-1}\langle E_{sp}|L_y|E_{sp}\rangle - u_s^{-1}\langle E_s|L_y|E_s\rangle, \tag{1a}$$

$$\text{or}\quad \Delta\langle L_y\rangle = 2k_0' U_{sp}^{-1}\int d^3\mathbf{r}'\,[(\mathbf{r}'-\mathbf{r}_{sp}')\times(\mathbf{E}_{sp}\times\mathbf{H}_{sp}^*)]_y - 2k_0 U_s^{-1}\int d^3\mathbf{r}'\,[(\mathbf{r}'-\mathbf{r}_s')\times(\mathbf{E}_s\times\mathbf{H}_s^*)]_y, \tag{1b}$$

provided that the correct $L_y$ operator is used in Eq. (1a) and the origin is the spacetime centre of energy. In Eq. (1b), $\mathbf{r}'$ refers to spacetime coordinates in the group velocity frame of the pulse (see below), $u_{s,sp} = \int d^3\mathbf{r}'\,|\mathbf{E}_{s,sp}|^2$, $U_{s,sp} = \int d^3\mathbf{r}'\,(|\mathbf{E}_{s,sp}|^2 + |\mathbf{H}_{s,sp}|^2)$, and $\mathbf{r}_{s,sp}' = U_{s,sp}^{-1}\int d^3\mathbf{r}'\,\mathbf{r}'(|\mathbf{E}_{s,sp}|^2 + |\mathbf{H}_{s,sp}|^2)$ are the spacetime centres of energy. Because STOV pulses are polychromatic [30], here $k_0$ is the central wavenumber and $k_0'$ allows for a central wavenumber shift in a spatiotemporally perturbed pulse. For weak perturbations, $k_0' = k_0$, and for negligibly absorbing or backscattering perturbations (see Sec. III), $U_{sp} = U_s$ (and $u_{sp} = u_s$). Our experimental perturbations are both weak and negligibly absorbing. The expressions in Eq. (1) also assume nonmagnetic material and $Re(n) \cong 1$, the conditions of our experiments.

As indicated, care must be taken in determining the form of the spatiotemporal OAM operator $L_y$. Unlike longitudinal OAM $L_z$, whose physical origin is the circulation of electromagnetic energy density around the z-axis in both $x$ and $y$ dimensions, energy density flow in a $\hat{\mathbf{z}}$-propagating STOV pulse in vacuum, with OAM along $\hat{\mathbf{y}}$, can occur only along $\pm x$: if any flow occurred along z, it would be superluminal or subluminal above or below the vortex singularity (depending on the sign of the STOV), violating special relativity. In recent work [30], we found a transverse spatiotemporal OAM operator, expressed in spacetime rectangular or polar coordinates,

$$L_y = (\mathbf{r}' \times \hat{\mathbf{p}}_{st})_y = -i\left(\xi \frac{\partial}{\partial x} + \beta_2 x \frac{\partial}{\partial \xi}\right)$$

$$= -i\left[\rho \sin \Phi \cos \Phi (1 + \beta_2)\frac{\partial}{\partial \rho} + (\cos^2 \Phi - \beta_2 \sin^2 \Phi)\frac{\partial}{\partial \Phi}\right] \quad (2)$$

$$\rightarrow -i(\cos^2 \Phi - \beta_2 \sin^2 \Phi)\frac{\partial}{\partial \Phi},$$

which applies in a dispersive optical material. Here $\xi = v_g t - z$ is a local space coordinate in the group velocity ($v_g$) frame of the pulse (local time is $\tau = \xi/v_g$), $t$ is time in the lab frame, $\beta_2 = v_g^2 k_0 k_0''$ is the dimensionless group velocity dispersion of the material, $k_0'' = (\partial v_g^{-1}/\partial \omega)_{k_0}$, and $\hat{\mathbf{p}}_{st} = -i\nabla_{st} = -i(\nabla_\perp - \beta_2 \partial/\partial \xi)$ is the spatiotemporal linear momentum operator [30]. The spacetime polar coordinates ($\rho, \Phi$) are defined by $x = \rho \sin \Phi$ and $\xi = \rho \cos \Phi$. The arrow in Eq. (2) indicates that the first term in the full polar coordinate expression always integrates to zero in Eq. 1(a), leaving the second term as the intrinsic transverse OAM operator in polar coordinates [30]. In performing the integrals in Eqs. (1a) and (1b), $d^3\mathbf{r}' = dydxd\xi$ in spacetime rectangular coordinates and $d^3\mathbf{r}' = dy\rho d\rho d\Phi$ in spacetime polar coordinates.

Our $L_y$ operator is consistent with special relativity, it conserves electromagnetic energy density flux, and it is conserved with propagation: $d/dz \langle L_y \rangle = i(2k_0)^{-1}\langle [H, L_y] \rangle = 0$ [30]. Here $[H, L_y] (= 0)$ is the commutator of $L_y$ and the propagation operator, $H = -\nabla_\perp^2 + \beta_2 \partial^2/\partial \xi^2$, from the spacetime paraxial wave equation $2ik_0 \partial \mathbf{A}(\mathbf{r}_\perp, \xi; z)/\partial z = H\mathbf{A}(\mathbf{r}_\perp, \xi; z)$ for the field $\mathbf{A}$. In the group velocity frame, $z$ plays the role of a time-like running parameter. It is straightforward to show that $L_y$ is conserved under non-paraxial propagation as well (see Appendix A).

For the room air of our experiments, $\beta_2 \cong 1.5 \times 10^{-5}$. This small dispersion has a negligible effect over short air propagation distances, so for the analysis below we take $\beta_2 = 0$ and the transverse OAM operator becomes $L_y = -i\xi \partial/\partial x \rightarrow -i \cos^2 \Phi \partial/\partial \Phi$ from Eq. (2). Note that without the gradient in $\xi$ enabled by nonzero $\beta_2$, this operator cannot transport energy density along $\pm \xi$ in the group velocity frame; it is transported only along $\pm x$.

To illustrate how STOVs propagate and to provide definitions for parameters used later in this paper, Fig. 1(a) reproduces results from [30], where a STOV pulse, of topological charge $l = 1$ and spacetime asymmetry ratio $\alpha = w_{0\xi}/w_{0x} = 0.24$, propagates right to left in air from $z/z_{0x} = -0.41$ to $z/z_{0x} = 0.24$, and right to left within each panel. Here $w_{0\xi}$ and $w_{0x}$ are time-like and space-like Gaussian spatial scales of the pulse, and $z_{0x} = k_0 w_{0x}^2/2$ is the space-like Rayleigh range. The top two rows are spatiotemporal intensity and phase profiles from the analytic modal STOV theory of [30], and the bottom rows are the corresponding experimental intensity and phase profiles $|E_s(x, \xi)|^2$ and $\phi_s(x, \xi) = \arg(E_s(x, \xi))$ captured by TG-SSSI [17], where the Gaussian $y$-dependence of the field is not displayed as it remains unaffected in our experiments and computations. Spatial diffraction along $\pm x$ causes the donut shape near $z = 0$ to evolve to lobed structures with opposite spacetime tilt on either side of $z = 0$, while the transverse OAM is conserved throughout propagation. For a STOV pulse of topological charge $l$ and spacetime asymmetry ratio $\alpha$ propagating in a dispersive medium characterized by $\beta_2$, the expectation value of transverse OAM is $\langle L_y \rangle = \frac{1}{2}l(\alpha - \beta_2/\alpha)$, so that for the air-propagating STOV pulse in Fig. 1, $\langle L_y \rangle = \frac{1}{2}l\alpha = 0.12$. The factor of ½ for the spacetime vortex is a direct result of energy density circulation restricted to $\pm x$ [30].

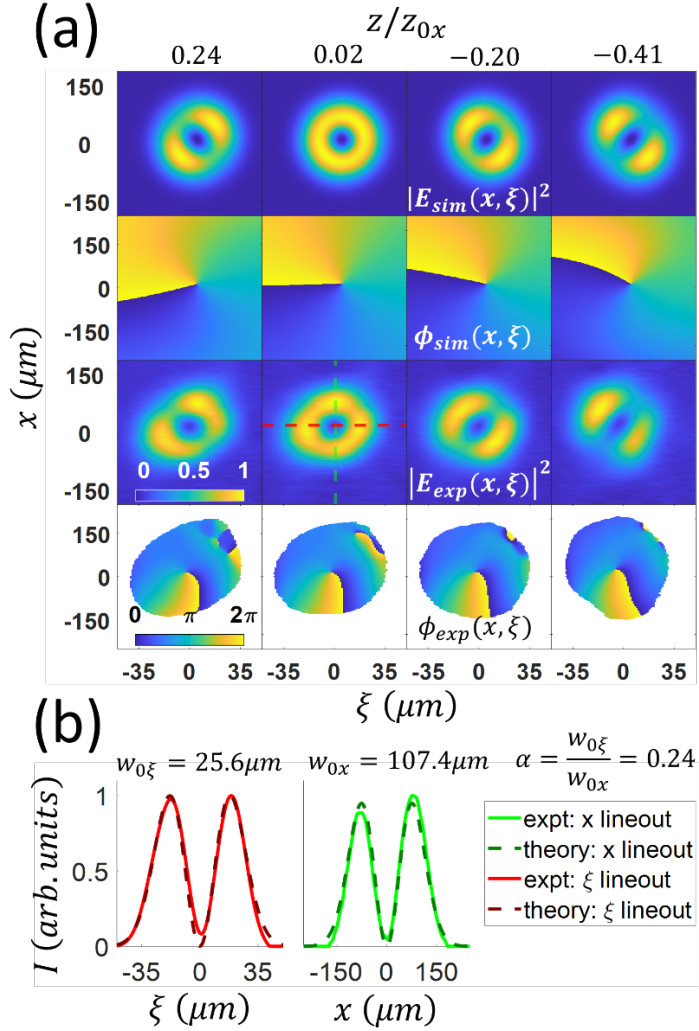

**Figure 1.** Illustrative figure on STOV pulse propagation for spacetime asymmetry ratio $\alpha = w_{0\xi}/w_{0x} = 0.24$, with $z_{0x} = \frac{1}{2}k_0 w_{0x}^2 = 4.5$ cm **(a)** *Top 2 rows*: Modal theory [30] plots of spatiotemporal intensity and phase of $l=1$ STOV pulse propagating right to left through its beam waist from $z/z_{0x} = -0.41$ to $z/z_{0x} = 0.24$ (and right to left within each panel). *Bottom 2 rows:* experimental intensity and phase plots extracted by TG-SSSI. **(b)** Lineouts along $(0,\xi)$ and $(x,0)$ of the experimental intensity profile at $z/z_{0x} = 0.02$ (solid lines) and fits to the modal theory curves (dashed lines).

### III. SPATIOTEMPORAL TORQUE

The goal of our experiments is twofold: (1) to explore how spatiotemporal perturbations to electromagnetic fields affect transverse OAM, and (2) to verify the correctness of our theoretical approach [30]. For an initial pulse $A_s(x,\xi) = |A_s(x,\xi)|e^{i\phi_s(x,\xi)}$ and a spatiotemporal perturbation $\Gamma(x,\xi) = |\Gamma(x,\xi)|e^{i\Delta\phi_p(x,\xi)}$, where $\phi_s(x,\xi)$ and $\Delta\phi_p(x,\xi)$ are real functions, the perturbed pulse is $A_{sp}(x,\xi) = \Gamma(x,\xi)A_s(x,\xi)$. This formulation implicitly assumes that the perturbation does not backscatter light back into the pulse; this condition is well satisfied by sufficiently weak perturbations, including those of our experiments, and by perturbations that effectively remove energy from the pulse. We take $A_s$ and $A_{sp}$ to be polarized along $\hat{y}$ so there are no effects of spin angular momentum. The change of transverse OAM per photon from the perturbation is then (see Appendix A)

$$\Delta\langle L_y\rangle = \langle L_y\rangle_{sp} - \langle L_y\rangle_s = iu_{sp}^{-1}\int dx\,d\xi\,\left[|A_s|^2|\Gamma|^2 L_y\Delta\phi_p + |A_s|^2\left(|\Gamma|^2 - \frac{u_{sp}}{u_s}\right)L_y\phi_s\right], \quad (3)$$

where $u_{sp} = \int dx\,d\xi\,|A_s(x,\xi)|^2|\Gamma(x,\xi)|^2$.

Equation (3) is intuitively appealing. The first term suggests the notion of "spatiotemporal torque", where the change in OAM is given by an effective force-lever arm product, $iL_y\Delta\phi_p = \xi\,\partial\Delta\phi_p/\partial x + \beta_2 x\,\partial\Delta\phi_p/\partial\xi$, weighted by the energy density distribution $|A_{sp}(x,\xi)|^2 = |\Gamma(x,\xi)A_s(x,\xi)|^2$ of the torqued object. Here the "force" components are $\partial\Delta\phi_p/\partial x$ and $\beta_2\,\partial\Delta\phi_p/\partial\xi$, and the lever arm components are $\xi$ and $x$. A mechanical analogy for the second term is the change in OAM caused by location-specific mass removal from a spinning wheel. For cases where energy is removed from the pulse by absorption or backscattering, $u_{sp}/u_s < 1$, and the second term contributes to the change in OAM provided that the initial pulse has transverse OAM; otherwise $L_y\phi_s = 0$ and the second term vanishes. That is, the wheel must already be spinning for mass removal to change OAM. Note that the second term will vanish, irrespective of $\phi_s$, in the case of a pure phase perturbation where $|\Gamma| = 1$ and $u_{sp}/u_s = 1$. This type of perturbation corresponds to our experiments.

Further examination of Eq. (3) leads to several insights: (a) Pure amplitude perturbations (with $\Delta\phi_p =$ constant) that conserve pulse energy cannot change the transverse OAM of a light pulse; in that case $\Gamma(x,\xi)$ can be viewed as a scattering coefficient that redistributes pulse energy over $\pm x$ in the pulse frame. (b) Steady state ($\partial\Delta\phi_p/\partial\xi = 0$) or spatially uniform ($\partial\Delta\phi_p/\partial x = 0$) phase perturbations do not change transverse OAM. (c) The only ways to change transverse OAM are ($i$) if either or both of the effective force terms, $\partial\Delta\phi_p(x,\xi)/\partial x$ and $\beta_2\,\partial\Delta\phi_p(x,\xi)/\partial\xi$, are time-varying and have an asymmetric temporal overlap with the energy density distribution (across the pulse's temporal centre of energy) or ($ii$) energy is removed from a pulse already containing transverse OAM. In atmospheric density gases, $\beta_2$ is negligible and $\partial\Delta\phi_p(x,\xi)/\partial x$ is the dominant contribution to the first term of Eq. (3), providing an unbalanced $x$-component force across $\xi = 0$.

We now consider a simple step function perturbation model, which provides good physical intuition and corresponds to our experiments (see Sec. IV). It also provides interpretative insight for the first experiment to measure STOVs [11]. We apply the spacetime perturbation $\Gamma(x,\xi) = |\Gamma(x,\xi)|e^{i\Delta\phi_p(x,\xi)}$, with $|\Gamma(x,\xi)| = 1$ and

$$\Delta\phi_p(x,\xi) = \Delta\phi_{p0}[\Theta(x - x_0 + h) - \Theta(x - x_0 - h)]\Theta(\xi - \xi_0), \quad (4)$$

to either a Gaussian pulse, $A_G(x,\xi)$, or to a $l = 1$ STOV pulse, $A_{STOV}(x,\xi)$:

$$A_G(x,\xi) = A_0\exp\left(-x^2/w_{0x}^2 - \xi^2/w_{0\xi}^2\right) \quad (5a)$$

$$A_{STOV}(x,\xi) = \left(\xi/w_{0\xi} + i\,x/w_{0x}\right)A_G(x,\xi) \quad (5b)$$

Here $\Theta(q)$ is the Heaviside function, $2h$ is the spatial width of the perturbation centered at $x = x_0$, and the perturbation turns on at $\xi = \xi_0$ (or $\tau = \tau_0$). The choice of a phase-only perturbation ($|\Gamma(x,\xi)| = 1$) corresponds to our experimental perturbation (see Sec. IV). The space-like and time-like widths of the Gaussian pulse are $w_{0x}$ and $w_{0\xi}$, and the expressions for $A_G$ and for $A_{STOV}$ are accurate for $z \ll z_{0x} = k_0 w_{0x}^2/2$ [30]. Because transverse OAM is conserved with $z$, it is sufficient to use these expressions in our calculations. Using Eq. 1(a) with initial fields $A_G$ or $A_{STOV}$, the simple perturbation model produces analytic solutions for $\Delta \langle L_y \rangle_G$ and $\Delta \langle L_y \rangle_{STOV}$ as a function of $(x_0, \xi_0)$:

$$\Delta \langle L_y \rangle_G = \frac{\Delta \phi_{p0}}{2\pi} \left( \alpha + \frac{\beta_2}{\alpha} \right) \left( \exp\left( \frac{8hx_0}{w_{0x}^2} \right) - 1 \right) \exp\left( -\frac{2(h+x_0)^2}{w_{0x}^2} - \frac{2\xi_0^2}{w_{0\xi}^2} \right) \tag{6a}$$

$$\begin{aligned}
\Delta \langle L_y \rangle_{STOV} &= \frac{\Delta \phi_{p0}}{2\pi} \left( \alpha + \frac{\beta_2}{\alpha} \right) \exp\left( -\frac{2(h+x_0)^2}{w_{0x}^2} - \frac{2\xi_0^2}{w_{0\xi}^2} \right) \\
&\times \left[ \left( \exp\left( \frac{8hx_0}{w_{0x}^2} \right) - 1 \right) \left( 1 + 2\frac{h^2 + x_0^2}{w_{0x}^2} + 2\frac{\xi_0^2}{w_{0\xi}^2} \right) - \frac{4hx_0}{w_{0x}^2} \left( \exp\left( \frac{8hx_0}{w_{0x}^2} \right) + 1 \right) \right].
\end{aligned} \tag{6b}$$

Figures 2(a) and 2(b) plot $\Delta \langle L_y \rangle_G$ and $\Delta \langle L_y \rangle_{STOV}$ vs. $(x_0, \xi_0)$. Each panel is for a particular half-width $h/w_{ox}$ and normalized dispersion $\beta_2$, where $\beta_2 = \pm 1$ is for dense, positively or negatively dispersive media and $\beta_2 = 0$ corresponds to low density media such as air, the propagation medium of our experiment. Plots using the dispersion of air, $\beta_2 \cong 1.5 \times 10^{-5}$, are indistinguishable from those using $\beta_2 = 0$.

We first discuss the $\beta_2 = 0$ plots (the plots for $\beta_2 = 1$ cases are qualitatively similar). Figure 2(a) shows the transfer of transverse OAM to $A_G$, a pulse with zero initial OAM. Maximum OAM transfer occurs for $x_0$ located at the spatial edges of the pulse ($x_0 \sim \pm w_{0x}$) and for $\xi_0$ located near the pulse centre ($\xi_0 \sim 0$). As discussed earlier in the context of Eq. (3), these optimum zones of $(x_0, \xi_0)$ maximize the overlap of the force-lever arm product with the torqued pulse energy density. Importantly, the perturbation transient (here, the step $\Theta(\xi - \xi_0)$) must overlap with the pulse so that the torque contributions are imbalanced across the temporal centre of energy at $\xi = 0$. For $\Theta(\xi - \xi_0)$ located outside the region of the pulse (for $|\xi_0/w_{0\xi}| > \sim 1$), $\Delta \langle L_y \rangle \to 0$ because the pulse sees the perturbation as steady state. The effect on $\Delta \langle L_y \rangle$ of a spatiotemporal torque localized in space and time is described in Appendix A, supporting the analogy of mechanical torque on a wheel.

The effect of the perturbation on an $l = 1$ STOV pulse is plotted in Fig. 2(b). Based on our prior discussion, it is not surprising that the plots are qualitatively similar to those for $l = 0$ in Fig. 2(a), with similar values of maximum OAM transfer $|\Delta \langle L_y \rangle|_{max}$. For a phase perturbation with $|\Gamma(x,\xi)| = 1$, Eq. (3) shows that $\Delta \langle L_y \rangle$ does not depend on $\phi_s(x,\xi)$, the phase winding of the initial pulse. Detailed differences between Figs. 2(a) and 2(b) arise from the different energy density distributions $|A_s(x,\xi)|^2$ for Gaussian and STOV pulses.

For $\beta_2 = -1$, $\Delta \langle L_y \rangle = 0$ in all cases. In such a negatively dispersive material, the spatiotemporal pulse shape is preserved because the dispersion in time matches the amplitude and sign of diffraction in space. The effect of any spatiotemporal torque applied to the pulse is zero, because the effective forces applied at the end of the lever arm are balanced.

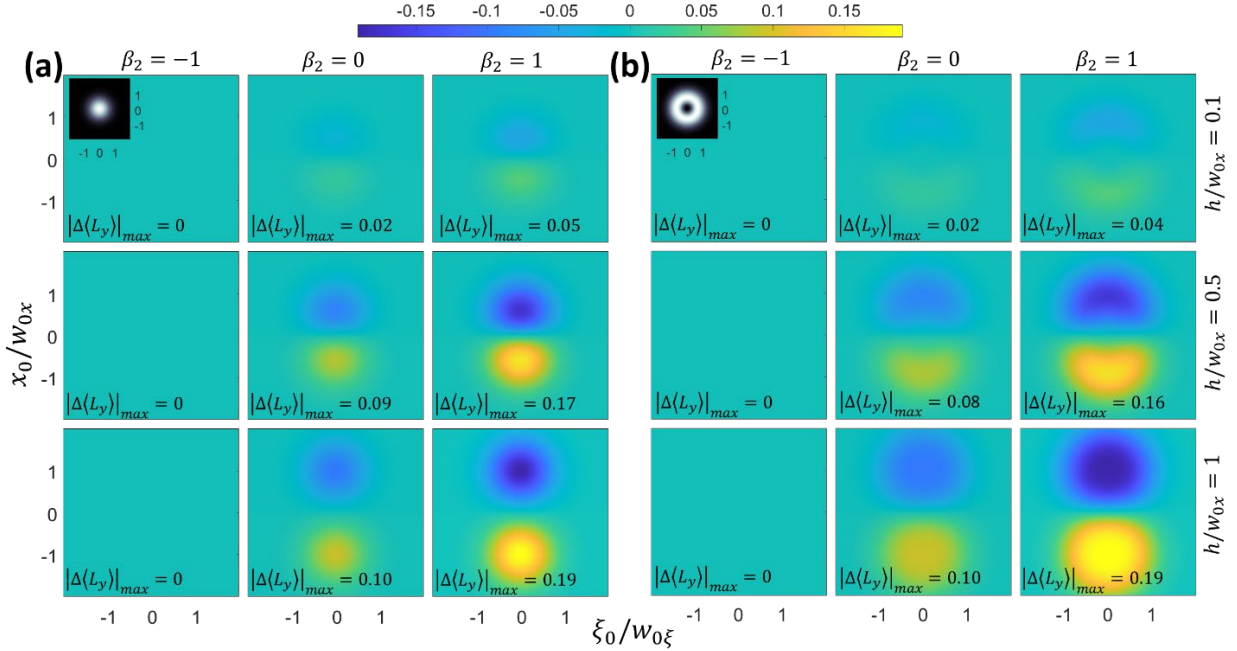

**Figure 2.** Plots of analytic solutions (Eqs. 6(a) and 6(b)) of $\Delta\langle L_y\rangle$ vs. $(x_0,\xi_0)$ for the spatiotemporal phase shift $\Delta\phi_p(x,\xi)$ of Eq. (4) applied to **(a)** a Gaussian pulse $A_G(x,\xi)$ (Eq. 5(a)) and to **(b)** an $l=1$ STOV pulse $A_{STOV}(x,\xi)$ (Eq. (5b)); $x_0$ and $\xi_0$ are the central space location and turn-on time of the perturbation. In Eq. (4), we choose $\Delta\phi_{p0}=-0.5$ to model the plasma generated by optical field ionization (OFI) of air.

It is important here to relate these results to the first experimental measurement of STOVs, which were spontaneously generated as a consequence of arrested self-focusing collapse in femtosecond pulse filamentation in air [11]. In air, femtosecond filamentation [34,35] occurs when an ultrashort pulse undergoes self-focusing collapse, which continues and accelerates until the intensity is high enough to ionize air molecules via optical field ionization (OFI), with the ultrafast-risetime plasma then acting to defocus the pulse. The few-femtosecond risetime of the plasma, determined by the OFI rate, occurs well within the pulse temporal envelope. The generated plasma then has a recombination-limited lifetime of several nanoseconds, an extremely long timescale compared to the pulse itself. The phase perturbation imparted by this plasma is therefore quite well modeled by Eq. (4) for $(x_0,\xi_0)=(0,0)$ and $h/w_{0x}\sim 0.5$, where the filament plasma is centered on the pulse and its width $2h$ is narrower than the beam width $\sim 2w_{0x}$.

While the filament-like case of $\beta_2=0$ and $h/w_{0x}=0.5$ (middle right panel in Fig. 2(a)) shows that $\Delta\langle L_y\rangle=0$ for $(x_0,\xi_0)=(0,0)$, the change in transverse OAM *density*, $\Delta M_y(x,\xi)=A_{sp}^*L_yA_{sp}$ (from Eq. 1(a)), is non-zero, and this is the effect measured in the experiment of ref. [11], albeit for a much larger $|\Delta\phi_{p0}|$ accumulated over self-focused propagation in ionizing air. The plots of $\Delta M_y$ vs. $(x,\xi)$ in Fig. 3(a) and (b) show regions of OAM density of opposite sign across the $x=0$ axis, displaying physics similar to an $x$-$\xi$ planar slice of the toroidal STOVs first measured in [11]. Figure 3(a) plots $\Delta M_y$ for the simple step function perturbation of Eq. (4); non-zero $\Delta M_y$ regions are only 1 pixel wide. Figure 3(b) uses a more realistic perturbation with smoothed step transitions,

$$\Delta\phi_p(x,\xi) = \tfrac{1}{2}\Delta\phi_{p0}\left(1 + \text{erf}\left(\sqrt{2}\,(\xi-\xi_0)/h_\xi\right)\right)\exp\left(-((x-x_0)/h_x)^8\right), \quad (7)$$

where we take $(x_0,\xi_0) = (0,0)$, $h_x/w_{0x} = 0.5$, and $h_\xi/w_{0\xi} = 0.5$.

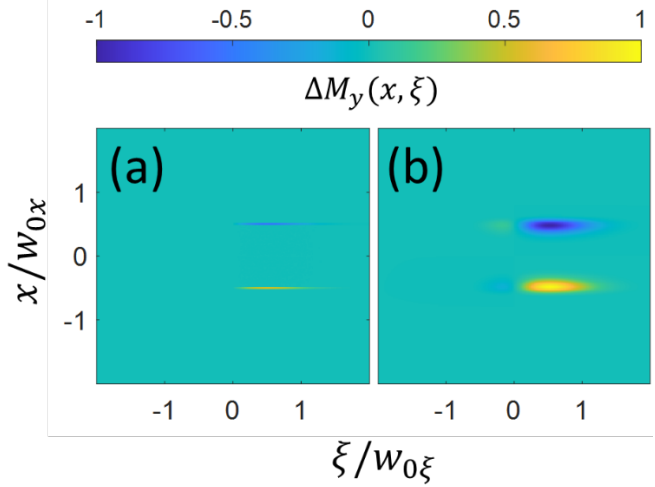

**Figure 3. (a)** Change in transverse angular momentum density $\Delta M_y(x,\xi) = A_{sp}^* L_y A_{sp}$ of a Gaussian pulse $A_G(x,\xi)$ using step function perturbation, Eq. (4). **(b)** Same as (a) except using smoothed perturbation, Eq. (7). In both panels, $\Delta M_y(x,\xi)$ is normalized by the maximum value $|\Delta M_y(x,\xi)|_{max}$.

We now address the effect of a non-energy-conserving pure amplitude perturbation on pulses with and without initial transverse OAM. We place the perturbation $\Gamma(x,\xi) = 1 - \exp(-(x/h)^8)$ at the beam waist ($z=0$) of $l=1$ STOV and Gaussian pulses described by Eqs. (5a) and (5b), with $w_{0x} = w_{0\xi} = 100$ μm, and $2h = 100$ μm. This models a steady state obstruction in the pulse propagation path such as a solid wire of diameter $2h$ centered at $x=0$, which would remove pulse energy by a combination of backscattering and absorption. Shown in Fig. 4(a) are the unperturbed spatiotemporal intensity profiles $I_s(x,\xi) = |E_s(x,\xi)|^2$ for the STOV and Gaussian pulses at $z = 0^-$, followed by the perturbed pulses $I_{sp}(x,\xi) = |E_{sp}(x,y=0,\xi;z)|^2$ propagating from $z=0$ to $z=2z_{0x}$. These were computed by forward-propagating the electric and magnetic fields $\mathbf{E}_{sp}(x,y,\xi;z)$ and $\mathbf{H}_{sp}(x,y,\xi;z)$ from $z=0$ using our unidirectional pulse propagation code YAPPE (see Appendix C). The z-dependent change in transverse OAM $\Delta\langle L_y\rangle_z$ was calculated directly from the fields using Eq. (1b) and plotted in Fig. 4(b) as points every $0.1 z_{0x}$. In both cases, as expected, $\Delta\langle L_y\rangle_z$ remains constant after the perturbation owing to the conservation of $L_y$. It is seen that only the STOV pulse has its transverse OAM per photon changed. This is predicted by Eq. (3): the second term contributes only if $L_y\phi_s \neq 0$ (the first term in Eq. (3) is zero because this is a pure amplitude perturbation). Note that even though the perturbation is on the beam axis at $x=0$, $\Delta\langle L_y\rangle$ is still non-zero because energy is removed from the pulse at a specific location; this imposes a new spatiotemporal distribution of the remaining energy and thus a new transverse OAM per photon. Changing the x-position of the wire will change $\Delta\langle L_y\rangle$ through new spatiotemporal distributions of the remaining energy. The constant solid line overlaid on the points is determined by a calculation of $\Delta\langle L_y\rangle_{z=0}$ using the matrix elements of Eq. (1a) and agrees with the direct field calculation.

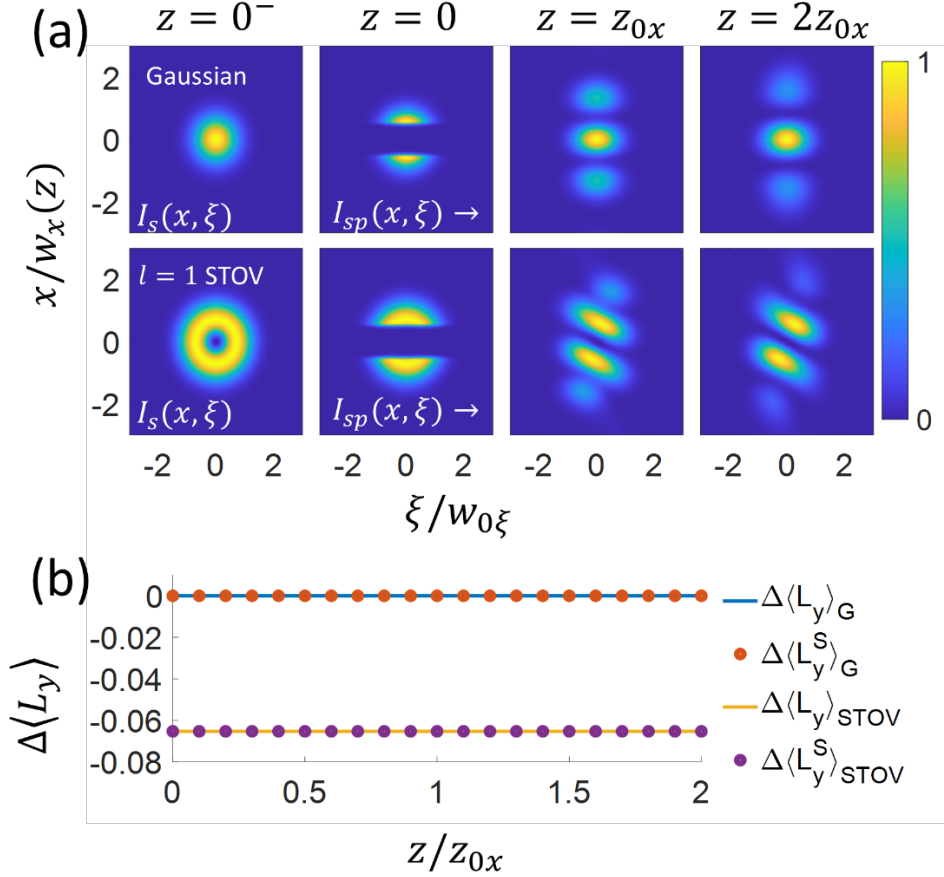

**Figure 4.** Effect of a non-energy-conserving pure amplitude perturbation $\Gamma(x,\xi) = 1 - \exp(-(x/h)^8)$ on pulses with and without transverse OAM. **(a)** Pre-perturbation $l=1$ STOV and Gaussian pulse intensities $|E_s(x,\xi)|^2$ at $z=0^-$, followed by the pulse intensity evolution $|E_{sp}(x,y=0,\xi;z)|^2$ from $z=0$ to $z=2z_{0x}$ determined by **E** and **H** field propagation computed by YAPPE (Appendix C). Here $2h=100$ μm and $w_{0x}=w_{0\xi}=100$ μm. **(b)** Change in transverse OAM per photon vs. $z$ ($\Delta\langle L_y\rangle_z$) for the Gaussian and STOV pulses calculated directly from the fields using Eq. 1(b) (points), and calculated using Eq. 1(a) (solid lines).

To conclude this section, it is important to make a connection to the generation of transverse OAM-carrying pulses using our $4f$ pulse shaper [14-16]. The shaper generates STOVs from zero-OAM Gaussian input pulses by applying torque in the spatio-spectral domain. One realization of the pulse shaper has a π-step phase plate in its Fourier $((x,\omega))$ plane and generates donut-shaped STOV pulses in the near field [16]. The phase jump in the $(x,\omega)$-plane, $\Delta\varphi_p(x,\omega) = \arg(\tilde{E}_{sp}(x,\omega))$, where $\tilde{E}_{sp}$ is the time Fourier transform of the shaper-perturbed pulse, plays a role analogous to the phase change $\Delta\phi_p(x,\xi)$ in the spatiotemporal domain. Because we exist in a spatiotemporal rather than a spatio-spectral world, with clocks marking time as the dynamical running parameter, it is *spatiotemporal* perturbations that naturally appear in physical phenomena occurring outside of carefully designed instruments such as $4f$ pulse shapers.

## IV. EXPERIMENTAL SETUP

The physical insight provided by the calculations of Sec. III led to our experimental design. To impart a spatiotemporal torque on an optical pulse via a perturbation $\Gamma(x,\xi) = |\Gamma(x,\xi)|e^{i\Delta\phi_p(x,\xi)}$, we impose a spatiotemporal refractive index perturbation in the propagation medium. This is accomplished by using a separate pulse to generate an ultrafast optical field ionization (OFI) air plasma at a controllable spacetime location; we call this spatiotemporal structure a "transient wire". As borne out by measurements and propagation simulations, the low-density plasma in the transient wire is dominantly a phase perturbation, with negligible energy removed from the pulse; effectively $|\Gamma(x,\xi)| = 1$. The transient wire has an ultrafast risetime and a narrow spatial width governed by the OFI rate, and a long lifetime governed by nanosecond timescale recombination: the spatiotemporal phase shift, $\Delta\phi_p(x,\xi)$, is therefore very well described by Eq. (4) or Eq. (7).

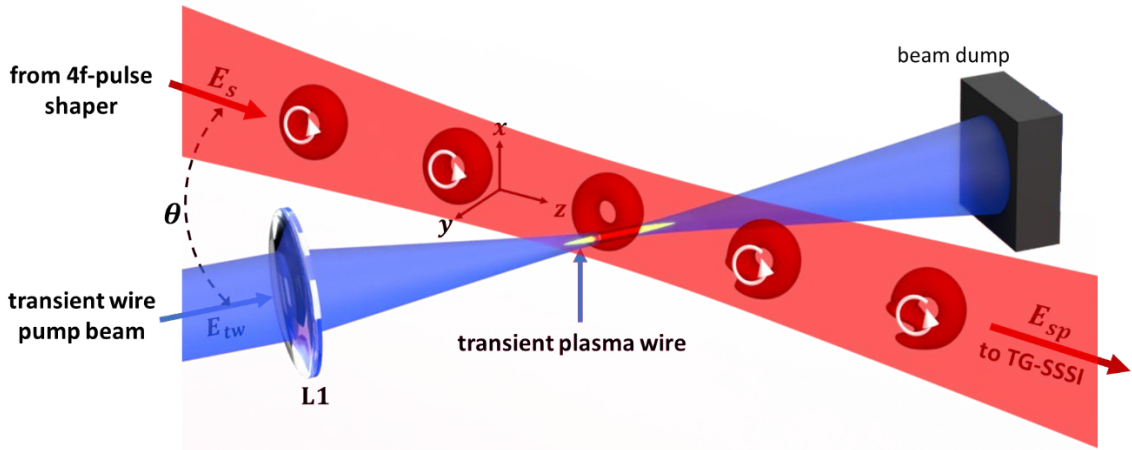

**Figure 5.** Configuration for measuring the effect of a transient phase perturbation on field $E_s$ (from a $4f$ pulse-shaper), imposed by the ultrafast optical-field-induced (OFI) plasma induced by field $E_{tw}$. This OFI plasma is the "transient wire". The perturbed pulse $E_{sp}$ and unperturbed pulse $E_s$ ($E_{tw}$ off) are measured by TG-SSSI (transient grating single shot supercontinuum spectral interferometry [16]). The angle between the beams is $\theta = 18.5°$. A detailed experimental diagram is shown in Appendix B.

Figure 5 is a schematic diagram of the transient wire experiment; a more detailed diagram is presented in Appendix B. Pulse $E_s$ (red beam), either a Gaussian or $l = 1$ STOV pulse from a $4f$ pulse shaper [14-17], propagates through air and is intersected by a focused secondary pulse $E_{tw}$ (blue beam) which generates an ultrafast risetime OFI plasma—the transient wire—at an adjustable spacetime location with respect to $E_s$. After the interaction, the perturbed pulse $E_{sp}$ is relay imaged from 3 mm past the interaction plane (to avoid nonlinear distortion in the imaging) to our TG-SSSI diagnostic [17], which extracts its spatiotemporal amplitude and phase. With $E_{tw}$ turned off, TG-SSSI measures the spatiotemporal amplitude and phase of the unperturbed pulse $E_s$. Five synchronized beams are needed for this experiment, which are obtained by splitting the output beam of a 1 kHz repetition rate Ti:Sapphire laser ($\lambda_0 = 800$nm, 40 fs) to give (1) an input pulse to the $4f$ pulse shaper, with output pulse $E_s$ (9.5 µJ, variable pulsewidth), (2) a focused transient wire beam $E_{tw}$ (~250 µJ, 40 fs FWHM, spot size $w_{tw} = 40$ µm) that intersects the $E_s$ beam at $\theta = 18.5°$, and (3) three pulses for TG-SSSI: twin probe and reference supercontinuum (SC) pulses $E_{pr}$ and $E_{ref}$ (with bandwidth $\Delta\lambda_{sc} \sim 160$ nm centred at $\lambda_{sc} = 630$ nm) plus a spatial interferometry reference pulse $\mathcal{E}_i$ (5.5 µJ after 3nm bandpass filter centered at 800nm). The angle $\theta = 18.5°$ is chosen to allow angular separation of the beams to direct $E_{sp}$ to the TG-SSSI diagnostic, and for sufficient spatial overlap of $\Delta\phi_p(x,\xi)$ along the propagation path of $E_s$.

## V. RESULTS AND DISCUSSION

An example of the transient wire perturbation of an $l = 1$ STOV pulse $E_s(x,\tau)$ is shown in Fig. 6, where here we use the local time coordinate $\tau = \xi/v_g$, and $(x,\tau) = (0,0)$ is taken as the spatiotemporal energy centre of $E_s$. Figure panels 5(a)-(c) show, respectively, the unperturbed $l = 1$ STOV pulse intensity $|E_s|^2$ ($E_{tw}$ off), the perturbed pulse intensity $|E_{sp}|^2$ ($E_{tw}$ on), and the transient-wire-induced phase shift $\Delta\phi_p(x,\tau) = \arg(E_{sp}) - \arg(E_s)$, all extracted using TG-SSSI [16,17]. The overlaid dashed red line shows the $x$-location of the perturbation, which was placed near the top of $E_s$ (at $x_0 = 120$ µm) to obtain appreciable $\Delta\langle L_y\rangle$, as motivated by the simulations in Fig. 2(b). The plots represent a $\Delta y \sim 10$ µm slice of $E_s$ and $E_{sp}$ in the $y$ direction, normal to the $x$-$\tau$ plane of the plots, where $\Delta y$ is the width of the imaging spectrometer slit used in TG-SSSI (see [17] and Appendix B).

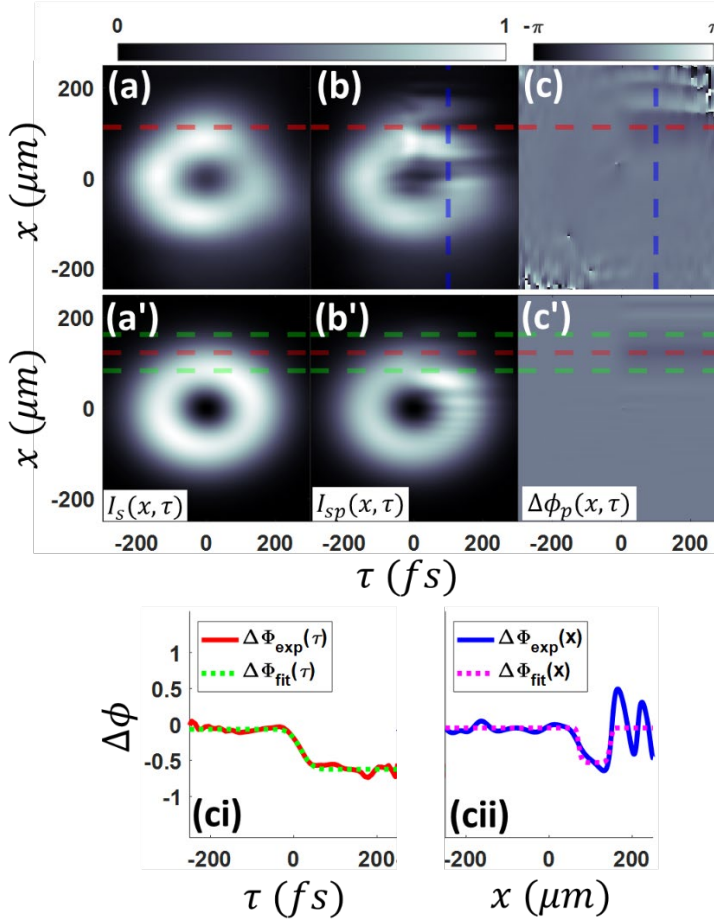

**Figure 6.** (a) TG-SSSI-measured $I_s = |E_s(x,\tau)|^2$ (transient wire off). (b) and (c) TG-SSSI-measured $|E_{sp}(x,\tau)|^2$ and $\Delta\phi_p(x,\tau)$ (transient wire on). (a′)-(c′) corresponding simulated $I_s^{sim} = |E_s^{sim}(x,\tau)|^2$, $I_s^{sim} = |E_{sp}^{sim}(x,\tau)|^2$, and $\Delta\phi_p^{sim}(x,\tau)$. **(ci)** Lineout of (c) along dashed red line (solid red) and fit (dotted green). **(cii)** Lineout of (c) along dashed blue line (solid blue) and fit (dotted pink). The fit curve neglects the oscillations on the right, which are due to the imaging plane being 3 mm past the interaction (see text). The fits in (ci) and (cii) are to $\Delta\phi_p^{sim}(x,\tau) = \frac{1}{2}\Delta\phi_{p0}(1 + \mathrm{erf}(\sqrt{2}(\tau - \tau_0)/h_\tau))\exp(-((x-x_0)/h_x)^8)$, giving $h_\tau = h_\xi/v_g = 44$ fs and $h_x = 40$ µm. In (ci) $x = 120$ µm ($= x_0$), and in (cii) $\tau = 100$ fs.

From Fig. 6(c), the maximum phase shift induced by the OFI plasma is $\Delta\phi_{p0} = -0.45$, where $E_{tw}$ was delayed so that the half-maximum phase shift $\Delta\phi_{p0}/2$, which defines the perturbation onset time $\tau_0$, occurred for $\tau_0 = 0$. From nonlinear least squares fitting of measured $\Delta\phi_p(x,\tau)$ to Eq. (7), we extract the phase shift risetime $h_\tau = h_\xi/v_g \sim 44$ fs and spatial half width at $1/e$ maximum $h_x \sim 40$ μm, with data lineouts overlaid with fits in Figs. 6(ci) and (cii). The peak phase shift corresponds to an OFI plasma density $\Delta N_e = |\Delta\phi_{p0}| N_{cr}\lambda_0/2\pi L \approx 5 \times 10^{17}$ cm$^{-3}$, where $N_{cr} = 1.7 \times 10^{21}$ cm$^{-3}$ is the critical density at $\lambda_0 = 800$ nm and $L = 2w_{tw}/\sin\theta \sim 250$ μm is the OFI plasma length experienced by $E_s$. It is seen in Fig. 6(b) that an amplitude modulation feature lies below the dashed red line, starting near $\tau = 0$. This modulation is the diffractive consequence of the OFI-induced phase perturbation, and develops during the 3 mm of propagation from the interaction location to the TG-SSSI object plane. This is borne out by the simulations of Fig. 6(a')-(c') (performed using YAPPE (Appendix C)), which show that similar diffractive modulations occur equidistantly above and below the dashed red line, but have no effect on the change in angular momentum of the pulse. Panel 6(a') shows the simulated unperturbed pulse $|E_s^{sim}|^2$ ($E_{tw}$ off) and panel 6(b') shows the perturbed pulse $|E_{sp}^{sim}|^2$ ($E_{tw}$ on), both 3 mm past the intersection with $E_{tw}$. Here, the perturbation by $E_{tw}$ is simulated by imposing on $E_s^{sim}$ the perturbation $\Gamma(x,\tau) = |\Gamma(x,\tau)|e^{i\Delta\phi_p(x,\tau)}$, with $|\Gamma(x,\tau)| = 1$ and $\Delta\phi_p^{sim}(x,\tau)$ from Eq. (7), using $h_\tau$ and $h_x$ derived from the fit discussed above. The red and green horizontal dashed lines in panels 6(a')-(c') mark the centre ($x = x_0$) and $\pm h_x$ edges of $\Delta\phi_p^{sim}(x,\tau)$.

In our main experiment, the results of which are shown in Fig. 7, we varied the spatiotemporal torque on both STOV and Gaussian pulses by scanning the transient wire onset time $\tau_0$. For torquing the STOV pulse, we spatially placed the wire near the top and bottom edges of the pulse, $x_0 = \pm 120$ μm (Figs. 7(a) and 7(b)), and at $x_0 = 60$ μm for the Gaussian pulse (Fig. 7(c)). The onset time was scanned from -200fs to 800fs in steps of $\Delta\tau_0 \sim 66$ fs, and the TG-SSSI-extracted complex spatiotemporal fields $E_s(x,\tau)$ and $E_{sp}(x,\tau)$ were then used to determine $\Delta\langle L_y\rangle$ at each delay using Eq. 1(a) and Eq. 1(b), which we label as $\Delta\langle L_y\rangle_{1a}$ and $\Delta\langle L_y\rangle_{1b}$. In Eq. 1(a), we use the STOV OAM operator $L_y = -i\xi\,\partial/\partial x$ (for $\beta_2 = 0$), while in Eq. 1(b), the $\mathbf{H}_{s,sp}(x,\xi)$ fields are calculated as the 2D inverse Fourier transform of $\tilde{\mathbf{H}}_{s,sp}(\mathbf{k},\omega) = (c/\omega)\mathbf{k}\times\tilde{\mathbf{E}}_{s,sp}(\mathbf{k},\omega)$, where $\tilde{\mathbf{E}}_{s,sp}(\mathbf{k}-\mathbf{k}_0,\omega)$ is the discrete 2D Fourier transform of the measured $\mathbf{E}_{s,sp}(x,\xi)$ fields, and $\mathbf{k}_0 = k_0\hat{\mathbf{z}}$ is the pulse central wavenumber.

The top two rows in Figs. 7(a)-(c) plot $I_{sp}(x,\tau) = |\mathbf{E}_{sp}(x,\tau)|^2$ and $\Delta\phi_p(x,\tau) = \arg(E_{sp}(x,\tau)) - \arg(E_s(x,\tau))$, with all amplitude and phase data extracted from raw TG-SSSI frames averaged over 500-750 shots. The lower panels plot $\Delta\langle L_y\rangle_{1a}$ and $\Delta\langle L_y\rangle_{1b}$ versus transient wire onset delay. These plots are in excellent agreement, confirming that our expression for the transverse OAM operator $L_y$ (ref. [30] and Eq. (2)) is correct. Overlaid in Figs. 7(a)-(c) are curves for $\Delta\langle L_y\rangle_{theory}$, using Eqs. 6(a) and 6(b) for $\Delta\langle L_y\rangle_G$ and $\Delta\langle L_y\rangle_{STOV}$. Agreement with the experimental results is excellent. In each experiment in Fig. 7, the measured experimental parameters were slightly different. These are listed in the figure caption and were incorporated into the expressions for $\Delta\langle L_y\rangle_G$ and $\Delta\langle L_y\rangle_{STOV}$.

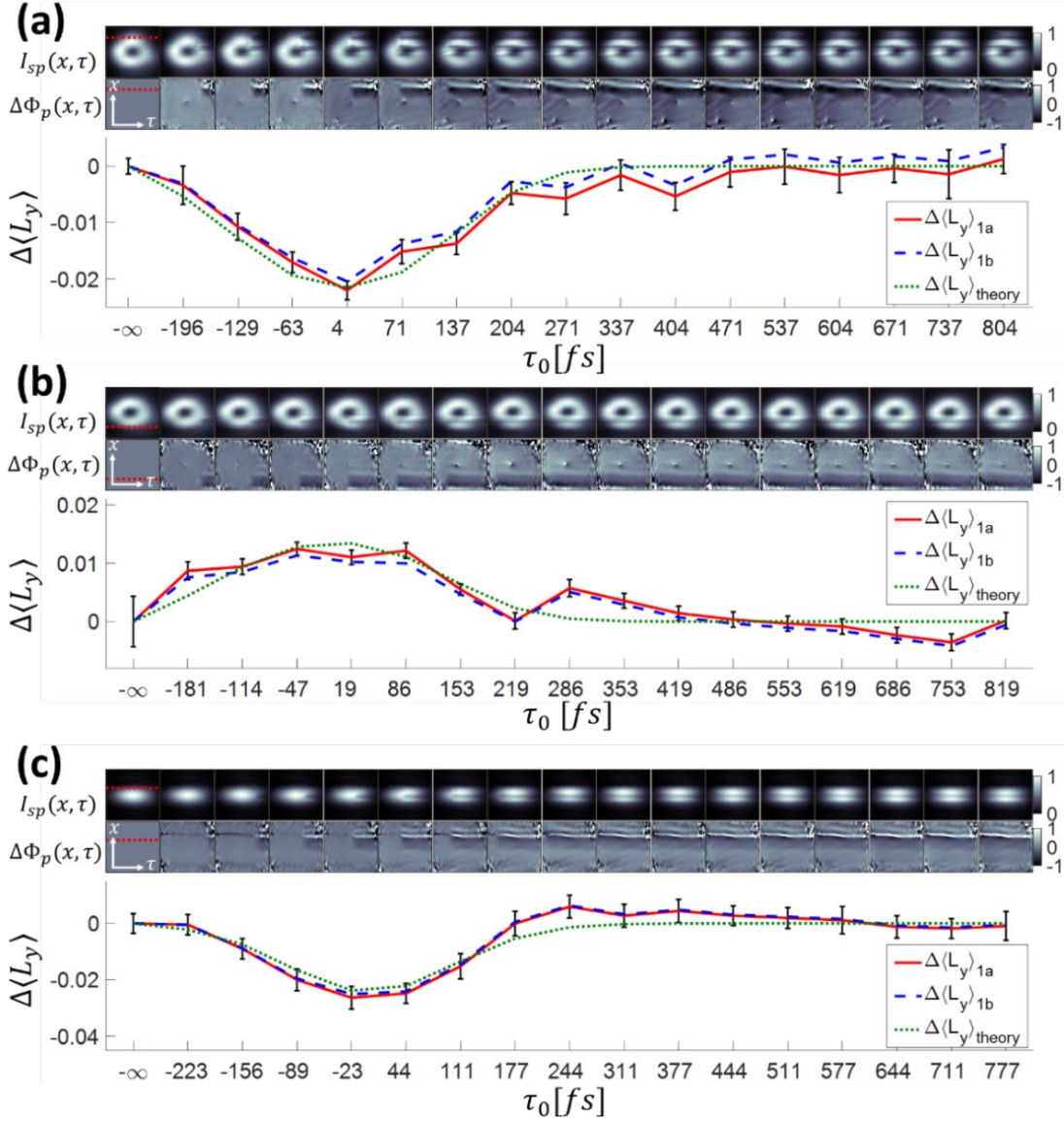

**Figure 7.** Effect of transient wire onset time $\tau_0$ on changing the transverse OAM of STOV and Gaussian pulses. Onset time is scanned between -200 fs and 800 fs in $\Delta\tau_0 = 66$ fs steps. **(a)** $l = 1$ STOV pulse, with transient wire centred at $x_0 = +120$ μm; **(b)** $l = 1$ STOV pulse, with transient wire centred at $x_0 = -120$ μm; **(c)** Gaussian pulse ($l = 0$), with transient wire centred at $x_0 = 60$ μm. The top two rows in (a)-(c) are $I_{sp}(x,\tau) = |E_{sp}(x,\tau)|^2$ and $\Delta\phi_p(x,\tau) = \arg(E_{sp}) - \arg(E_s)$, where $E_s(x,\tau)$ and $E_{sp}(x,\tau)$ are the unperturbed and perturbed complex fields extracted from TG-SSSI measurements. The spatial location of the perturbation is indicated by the red dotted lines. The bottom panels in (a)-(c) plot $\Delta\langle L_y\rangle_{1a}$ and $\Delta\langle L_y\rangle_{1b}$, which are the change in spatiotemporal OAM per photon computed by inserting the measured $E_s$ and $E_{sp}$ into Eq. 1(a), and the **E** and **H** fields computed from them into Eq. 1(b). The error bars are the ± standard deviation over 500-750 shots of extracted data. Overlaid in (a)-(c) is $\Delta\langle L_y\rangle_{theory}$, calculated using Eqs. (6a) and (6b), in which we use measured and fit quantities. For $\Delta\langle L_y\rangle_{theory}$ in panel (a): ($\Delta\phi_{p0} = -0.45, w_{0x} = 120$ μm , $w_{0\xi} = 56$ μm, , $h = 40$ μm, $x_0 = 120$ μm). For $\Delta\langle L_y\rangle_{theory}$ in panel (b): ($\Delta\phi_{p0} = -0.21, w_{0x} = 110$ μm , $w_{0\xi} = 56$ μm, $h = 40$ μm, $x_0 = -120$ μm). For $\Delta\langle L_y\rangle_{theory}$ in panel (c): ($\Delta\phi_{p0} = -0.31, w_{0x} = 100$ μm , $w_{0\xi} = 61$ μm, $h = 40$ μm, $x_0 = 60$ μm). The " $-\infty$" mark on the time axes refers to the $\xi_0 \to -\infty$ limit of $\Delta\langle L_y\rangle_{theory}$.

The results of Fig. 7 confirm our Sec. III theory: once the OFI plasma phase transient is shifted away from the pulse envelope, $\Delta\langle L_y\rangle \to 0$. To impart spatiotemporal torque and a change in transverse OAM, the perturbation transient must temporally overlap with the pulse energy density distribution. Refractive index transients with timescales much longer than the pulse temporal envelope have little effect on the transverse OAM of a pulse. In general, irrespective of its spatial location or peak amplitude, the more imbalanced a transient spatiotemporal perturbation is across the temporal centre of energy of an optical pulse, the greater effect it has on changing the transverse OAM.

## VI. CONCLUSIONS

We have demonstrated that the transverse orbital angular momentum (OAM) per photon of an electromagnetic pulse can be changed only by a transient phase perturbation comparable to the pulse envelope and overlapping with it, or by a non-energy-conserving amplitude perturbation if the pulse already has transverse OAM. Our half-integer theory of STOV pulse OAM [30] is in excellent agreement with our experiments and with propagation simulations directly using the **E** and **H** fields. The experiments of this paper, in which spatiotemporal torques were faster than the short pulses to which they were applied, would not have been possible without our high bandwidth, high time resolution single shot technique—TG-SSSI [17]— for STOV pulse amplitude and phase recovery.

The concept of spatiotemporal torque, introduced in this paper, provides insight into the dynamics leading to changes in transverse OAM: the effective force, manifested as a spatiotemporal phase gradient supplied by the perturbation, is weighted by the spacetime lever arm and the electromagnetic energy density distribution. If the initial field is a STOV pulse with zero energy density at the singularity (an "edge-first flying donut" [16]), spatiotemporal torquing can be analogized by mechanical torque on a rotating hoop, where maximum change in OAM is obtained by applying force at the outer rim, where the product of lever arm and mass density is maximum. However, unlike in the mechanical case, a spatiotemporal torque applied to an optical pulse changes the OAM of all particles (photons) identically. The other way to change transverse OAM is to remove energy from a pulse already with transverse OAM; this can be accomplished by a non-energy-conserving amplitude perturbation. This imposes a new spatiotemporal distribution of the remaining energy and thus a new transverse OAM per photon. Here, the mechanical analogy is location-specific mass removal from a spinning wheel.

Our results point the way to methods of distortion-free encoding of information in transverse OAM, for example, in propagation through turbulent atmosphere. The shortest transient timescale for turbulent refractive index fluctuations in the atmosphere is a few milliseconds [36], at least 10 orders of magnitude longer than a ~100 fs duration ultrashort pulse, so air turbulence acts as a weak stationary perturbation with no effect on the expectation value of transverse OAM per photon. While the turbulence-induced *spatial* phase shifts can manifest as transverse ($xy$) spatial distortion of the beam, the encoded spatio-temporal phase structure makes possible the extraction of time-based information with fast retrieval schemes.

## ACKNOWLEDGMENTS

The authors thank Nischal Tripathi for technical assistance. This work is supported by the Air Force Office of Scientific Research (FA9550-21-1-0405), Office of Naval Research (N00014-17-1-2705 and N00014-20-1-2233), and the National Science Foundation (PHY2010511).

## APPENDIX A: TRANSVERSE ORBITAL ANGULAR MOMENTUM OF LIGHT

### 1. Conservation of transverse OAM operator $L_y$ under non-paraxial propagation

In ref. [30], we showed that the transverse OAM operator $L_y = -i(\xi \, \partial/\partial x + \beta_2 x \, \partial/\partial \xi)$ was conserved under paraxial propagation. Here, we start with the non-paraxial propagation equation for the field envelope $A(\mathbf{r}_\perp, z, t)$

$$\frac{\partial^2 A}{\partial z^2} + i2k_0 \frac{\partial}{\partial z} = -\nabla_\perp^2 A - i2k_0 k_0' \frac{\partial A}{\partial t} + k_0 k_0'' \frac{\partial^2 A}{\partial t^2}, \tag{A1}$$

In the group velocity frame, using $\zeta = z$, $\xi = v_g t - z$, $H = -\nabla_\perp^2 + \beta_2 \, \partial^2/\partial \xi^2$, and $p_z^2 = -\partial^2/\partial z^2$ gives

$$\frac{\partial A}{\partial \zeta} = \frac{i}{2k_0}\left[HA - \left(\frac{\partial^2 A}{\partial \zeta^2} - 2\frac{\partial^2 A}{\partial \zeta \partial \xi} + \frac{\partial^2 A}{\partial \xi^2}\right)\right] = \frac{i}{2k_0}[H - p_z^2]A . \tag{A2}$$

Then for $\langle L_y \rangle = \langle A|L_y|A \rangle$,

$$\frac{d}{dz}\langle L_y \rangle = \left\langle \frac{\partial}{\partial z} A \middle| L_y \middle| A \right\rangle + \left\langle A \middle| \frac{\partial}{\partial z} L_y \middle| A \right\rangle + \left\langle A \middle| L_y \middle| \frac{\partial}{\partial z} A \right\rangle, \tag{A3}$$

Since $L_y$ does not explicitly depend on $z$, and since $H$, $L_y$, and $p_z$ are all Hermitian, Eq. (A3) becomes

$$\frac{d}{dz}\langle L_y \rangle = \frac{i}{2k_0}\langle A|[H, L_y]|A \rangle + \frac{i}{2k_0}\langle A|[L_y, p_z^2]|A \rangle = 0, \tag{A4}$$

because $L_y$ commutes with both $H$ and $p_z^2$.

### 2. Effect of a spatiotemporal perturbation on transverse OAM

The expectation values of $L_y$ per photon for the unperturbed and perturbed pulses, $A_s(x, \xi) = |A_s(x,\xi)|e^{i\phi_s(x,\xi)}$ and $A_{sp}(x,\xi) = \Gamma(x,\xi)A_s(x,\xi) = |\Gamma(x,\xi)|e^{i\Delta\phi_p(x,\xi)}A_s(x,\xi)$, are

$$\langle L_y \rangle_{s,sp} = u_{s,sp}^{-1}\langle E_{s,sp}|L_y|E_{s,sp}\rangle . \tag{A5}$$

This gives

$$\langle L_y \rangle_s = u_s^{-1} \int dx d\xi |A_s|^2 \left(\xi \frac{\partial \phi_s}{\partial x} + \beta_2 x \frac{\partial \phi_s}{\partial \xi}\right), \tag{A6a}$$

and

$$\langle L_y \rangle_{sp} = u_{sp}^{-1} \int dx\, d\xi\, \Big[-i|A_s|^2|\Gamma|\left(\xi \frac{\partial |\Gamma|}{\partial x} + \beta_2 x \frac{\partial |\Gamma|}{\partial \xi}\right)$$

$$- i|\Gamma|^2 |A_s|\left(\xi \frac{\partial |A_s|}{\partial x} + \beta_2 x \frac{\partial |A_s|}{\partial \xi}\right) + |A_s|^2|\Gamma|^2\left(\xi \frac{\partial \Delta\phi_p}{\partial x} + \beta_2 x \frac{\partial \Delta\phi_p}{\partial \xi}\right) \quad \text{(A6b)}$$

$$+ |A_s|^2|\Gamma|^2\left(\xi \frac{\partial \phi_s}{\partial x} + \beta_2 x \frac{\partial \phi_s}{\partial \xi}\right)\Big],$$

The first two terms in Eq. (A6b) integrate to zero, yielding

$$\Delta\langle L_y \rangle = \langle L_y \rangle_{sp} - \langle L_y \rangle_s$$
$$= iu_{sp}^{-1} \int dx\, d\xi\, \left[|A_s|^2|\Gamma|^2 L_y \Delta\phi_p + |A_s|^2\left(|\Gamma|^2 - \frac{u_{sp}}{u_s}\right)L_y\phi_s\right]. \quad \text{(A7)}$$

For an initial pulse with zero transverse OAM, such as a Gaussian, $L_y \phi_s = 0$ and

$$\Delta\langle L_y \rangle = iu_{sp}^{-1} \int dx\, d\xi\, |A_s|^2|\Gamma|^2 L_y \Delta\phi_p. \quad \text{(A8)}$$

For a phase-only perturbation $|\Gamma(x,\xi)| = 1$, $u_{sp} = u_s$, and Eq. (A8) also applies.

### 3. Assessment of an alternative transverse OAM operator

Recent work [28, 29] has asserted that the "canonical" operator for intrinsic transverse OAM is

$$\pounds_y = -i\left(\xi \frac{\partial}{\partial x} - x \frac{\partial}{\partial \xi}\right), \quad \text{(A9)}$$

where $x$ and $\xi$ are as defined earlier. In the formulation of [28, 29], $x$ and $\xi$ are treated on an equal footing, just as $x$ and $y$ are treated in the $L_z$ operator for longitudinal OAM.

Adoption of $\pounds_y$ assumes unphysical effects, including super- and sub-luminal energy density flow around the spatiotemporal vortex singularity in vacuum, and non-conservation [30,37]. While any valid angular momentum quantity should be conserved with propagation, $\pounds_y$ is not. Namely,

$$\frac{d}{dz}\langle \pounds_y \rangle = \frac{i}{2k_0}\langle [H, \pounds_y]\rangle = k_0^{-1}(1+\beta_2)\left\langle \frac{\partial^2}{\partial x \partial \xi}\right\rangle\bigg|_{z=0}, \quad \text{(A10)}$$

which is non-zero except if $\beta_2 = -1$ (when $L_y \to \pounds_y$) or when there is spatiotemporal field symmetry.

We test the consequences of using $\pounds_y$ with a spatiotemporally asymmetric field. Such a field can be generated, as seen in Figs. 6 and 7, when a spatiotemporal perturbation is applied to a symmetric pulse. Here, we apply at $z = 0^-$ a phase-only perturbation $\Gamma(x,\xi) = e^{i\Delta\phi_p(x,\xi)}$, with $\Delta\phi_p(x,\xi)$ from Eq. (7) (with $\Delta\phi_{p0} = -0.5$, $x_0 = -100$ μm, $\xi_0 = 0$, $h_x = 50$ μm, and $h_\xi = 50$ μm) to the Gaussian pulse $A_G$ of Eq. 5(a). The transverse OAM of $A_G$ is zero. Non-paraxial propagation evolution of the perturbed field $A_{sp}(x,y,\xi;z) = \Gamma(x,\xi)A_G(x,y,\xi;z)$ is then computed using our code YAPPE (Appendix C). In Fig. A1, we plot $\Delta\langle L_y\rangle_z = u_{sp}^{-1}\langle A_{sp}|L_y|A_{sp}\rangle_z$ and $\Delta\langle\pounds_y\rangle_z = u_{sp}^{-1}\langle A_{sp}|\pounds_y|A_{sp}\rangle_z$ as a function of propagation distance $z$. The immediate post-perturbation values, $\Delta\langle L_y\rangle_{z=0}$ and $\Delta\langle \pounds_y\rangle_{z=0}$, are shown in the figure; these differ. It is clear that $\Delta\langle L_y\rangle_z$ is conserved with propagation, while $\Delta\langle \pounds_y\rangle_z$ is not. The divergence of $\Delta\langle \pounds_y\rangle_z$ is predicted by Eq. (A10) and is a consequence of the non-commutation of $\pounds_y$ and the propagation operator $H$. This increase is linear in $z$, with slope $d\langle \pounds_y\rangle/dz = k_0^{-1}\langle \partial^2/\partial x\partial\xi\rangle_{z=0}$, and is in excellent agreement with $\Delta\langle \pounds_y\rangle_z = u_{sp}^{-1}\langle A_{sp}|\pounds_y|A_{sp}\rangle_z$. Calculating the change in transverse OAM using Eq. 1(b), with the **E** and **H** fields propagated non-paraxially by the YAPPE simulation, gives the blue points labelled as $\Delta\langle L_y^S\rangle_z$; these agree with $\Delta\langle L_y\rangle_z$.

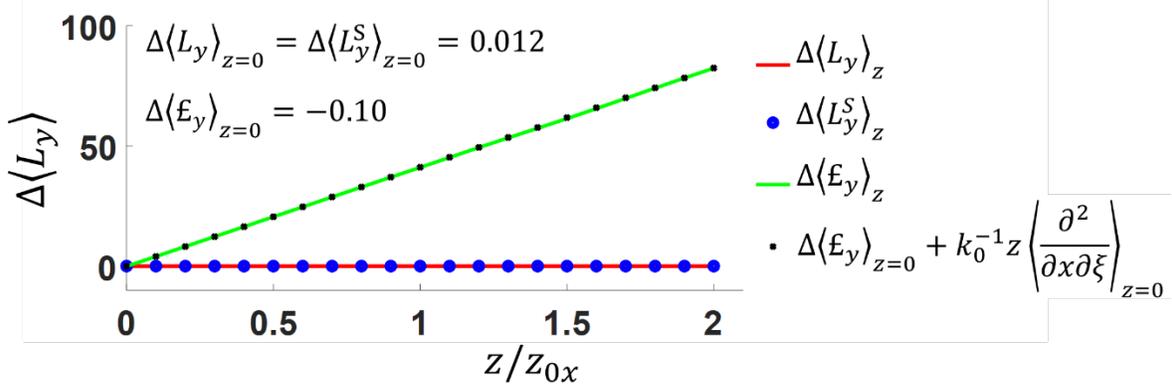

**Figure A1.** Propagation evolution of $\Delta\langle L_y\rangle = u_{sp}^{-1}\langle A_{sp}|L_y|A_{sp}\rangle_z$ (red curve) and $\Delta\langle \pounds_y\rangle = u_{sp}^{-1}\langle A_{sp}|\pounds_y|A_{sp}\rangle_z$ (green curve), where the z-evolution of $A_{sp}(x,y,\xi;z)$ is non-paraxially computed using our code YAPPE (Appendix C). The perturbation (given by Eq. (7), with $\Delta\phi_{p0} = -0.5$, $x_0 = -100$ μm, $\xi_0 = 0$, $h_x = 50$ μm, and $h_\xi = 50$ μm) is applied to the Gaussian pulse of Eq. 5(a), with $w_{ox} = 100$μm and $w_{0\xi} = 100$ μm. The immediate post-perturbation OAM changes are $\Delta\langle L_y\rangle_{z=0} = 0.012$ and $\Delta\langle \pounds_y\rangle_{z=0} = -0.10$. Also plotted: Large blue points $\Delta\langle L_y^S\rangle$ computed using Eq. 1(b), with the **E** and **H** fields propagated non-paraxially with the code YAPPE (Appendix C), and small black points computed as $\Delta\langle \pounds_y\rangle_z = \Delta\langle \pounds_y\rangle_{z=0} + k_0^{-1}z\langle \partial^2/\partial x\partial\xi\rangle_{z=0}$ from integration of Eq. (A10), where $\langle \partial^2/\partial x\partial\xi\rangle_{z=0} = \langle A_{sp}|\partial^2/\partial x\partial\xi|A_{sp}\rangle$ for $A_{sp} = A_{sp}(x,y,\xi;z=0)$.

Note that non-conservation of $\pounds_y$ is independent of choice of origin, whether it is the perturbed pulse centre of energy $\mathbf{r}'_{sp}$ (as in Fig. A1)) or the "photon number centroid" [29], so $\pounds_y$ cannot be corrected with extrinsic OAM contributions. This is because non-conservation of $\pounds_y$ is caused by the inclusion, even in vacuum, of a non-zero linear momentum density $p_\xi = -i\,\partial/\partial\xi$. As the propagating pulse spatially (transversely) diffracts, with its width in $x$ increasing, the contribution of $xp_\xi$ to $\pounds_y$ increases and $\langle A_{sp}|\pounds_y|A_{sp}\rangle_z$ unavoidably increases with propagation. There are circumstances, not involving pulse propagation, where $\pounds_y$ is appropriate to use. One such example is a vortex stationary in the lab frame, whose transverse OAM, say along $\hat{\mathbf{y}}$, can be described in $x$-$z$ space coordinates [38].

The choice of origin as the center of energy is essential to isolating the intrinsic OAM from the extrinsic OAM. While Eq. (A4) proves that OAM calculated for any choice of origin is conserved with propagation, to compute the intrinsic OAM in simulations and from experimental data, we always choose the center of energy as the origin [39], thereby automatically removing the extrinsic part of the OAM. A straightforward mechanical analogy also makes this point: the intrinsic OAM of a massive blob, consisting of point particles of varying mass, must be calculated with respect to the centre of mass density.

## 4. Application of a spatiotemporal phase perturbation localized in space and time

As discussed in Sec. IV, our transient wire spatiotemporal perturbation is well modeled by Eq. (4) or Eq. (7), which describe a narrow spatial structure with a fast turn-on time and no turn-off. To more finely map the effect of spatiotemporal perturbations on electromagnetic pulses, we consider a phase perturbation $\Gamma(x,\xi) = e^{i\Delta\phi_p(x,\xi)}$ localized in both space and time and centered at $(x_0, \xi_0)$: $\Delta\phi_p(x,\xi) = \Delta\phi_{p0} \exp[-(x-x_0)^2/h_x^2 - (\xi-\xi_0)^2/h_\xi^2]$. Figure A2 shows the change in transverse OAM per photon $\Delta\langle L_y\rangle_{x_0,\xi_0}$ for various pulses as a function of $(x_0, \xi_0)$, plotted from analytic expressions determined using Eq. (1a) [40]. In all cases, we take $\beta_2 = 0$ and the perturbation spatial width $h_x/w_{0x} = 0.25$, with the other parameters listed on the panels and in the figure caption.

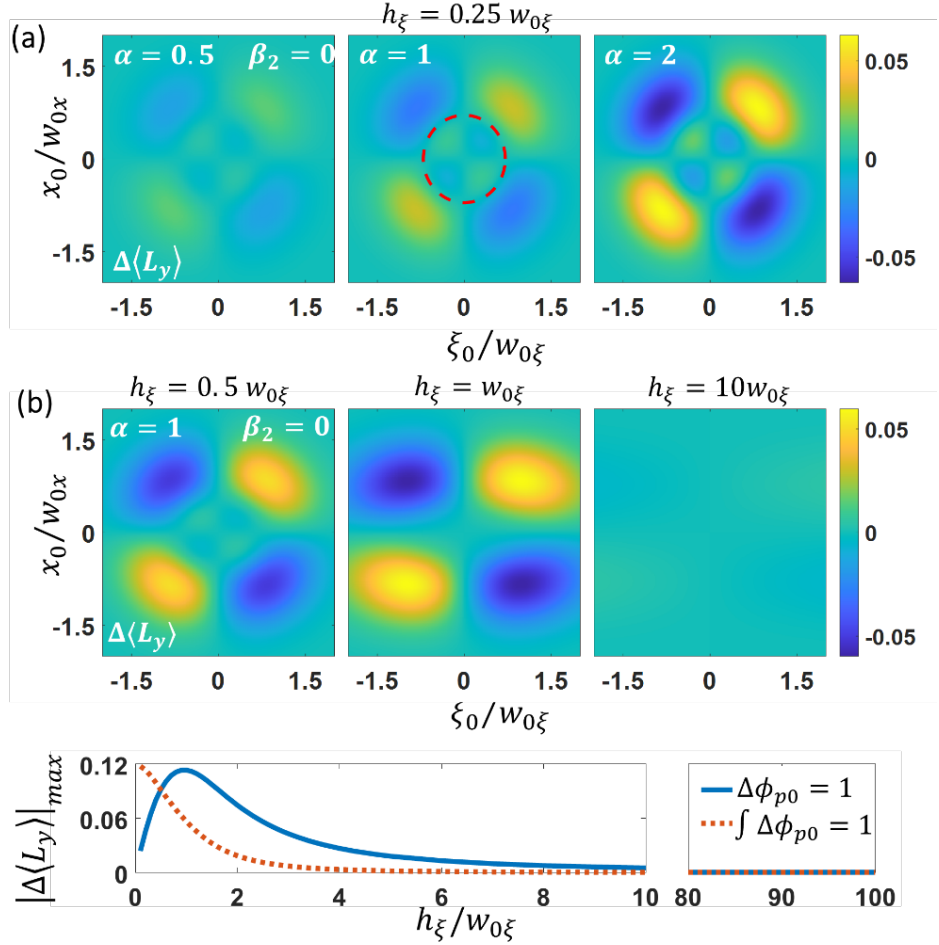

**Figure A2.** Plots of analytic solutions [40] for change in transverse OAM per photon, $\Delta\langle L_y\rangle_{x_0,\xi_0}$, imparted to an optical pulse as function of $(x_0,\xi_0)$ by a spatiotemporal phase perturbation $\Gamma(x,\xi) = \exp(i\Delta\phi_p(x,\xi))$. Here $\Delta\phi_p(x,\xi) = \Delta\phi_{p0}\exp[-(x-x_0)^2/h_x^2 - (\xi-\xi_0)^2/h_\xi^2$, $\Delta\phi_{p0} = 1$, $\beta_2 = 0$ and $h_x/w_{0x} = 0.25$. **(a)** perturbation $\Gamma(x,\xi)$ applied to a $l = 1$ STOV pulse $A_{\text{STOV}}(x,\xi) = (\xi/w_{0\xi} + i\,x/w_{0x})A_G(x,\xi)$ for $h_\xi/w_{0\xi} = 0.25$ and $\alpha = w_{0\xi}/w_{0x} = 0.5, 1,$ and $2$. The red dashed circle in the centre panel is the contour of peak intensity of $|A_{\text{STOV}}|^2$. **(b)** $\Gamma(x,\xi)$ applied to $A_{\text{STOV}}(x,\xi)$ as in (a), here with $\alpha = 1$ and transient width $h_\xi/w_{0\xi} = 0.5, 1,$ and $10$. The curves immediately below plot the maximum absolute change in OAM ($\left|\Delta\langle L_y\rangle_{x_0,\xi_0}\right|_{max}$) vs. transient width $h_\xi$, for the cases of fixed peak phase shift $\Delta\phi_{p0} = 1$ and constant integrated phase shift $(h_x h_\xi)^{-1}\int dx d\xi\,\Delta\phi_p(x,\xi) = 1$. The overlaid dashed ring follows the maximum intensity contour of $A_{\text{STOV}}$.

Examination of Fig. A2(a), for spatiotemporal torque applied to the $l = 1$ STOV pulse of Eq. 5(b), confirms our intuitive expectations from Sec. III. Torque is maximized when the perturbation peak $(x_0, \xi_0)$ is placed at the spatiotemporal locations with appreciable energy density and lever arm (see Eq. (3)), while dropping to zero when crossing lines marking the spatial and temporal centers of energy, where torque contributions from opposites sides in space and time cancel. Combined, the two effects give rise to the characteristic 4-lobed patterns plotted. A similar pattern appears in the torquing of Gaussian pulses [40]. It is interesting to note that the locations of maximum torque are spatiotemporally outside the peak intensity contour of $A_\text{STOV}$, which is marked with a dashed red circle. This is the effect of lever arm weighting of the optical energy density in Eq. (3). A weaker 4-lobed structure with opposite polarity can be seen inside the peak intensity contour; here $h_x/w_{0x} = 0.25$ and $h_\xi/w_{0\xi} = 0.25$ are small enough for the perturbation to torque the inside of the STOV "wheel". This structure disappears for the larger values of $h_\xi/w_{0\xi}$ in Fig. A2(b). Also evident is the linear scaling of $\Delta\langle L_y\rangle$ with $\alpha$, which stems from our theory of STOV transverse OAM [30].

The effect of a temporally widening $\Delta\phi_p(x,\xi)$ of fixed peak amplitude is plotted in Fig. A2(b). It is seen that in the middle panel (for the middle pulsewidth), the torque is both larger in size and is effectively applied over a wider spatiotemporal area than for perturbations of shorter and longer pulsewidths. In particular, the very long perturbation of the rightmost panel registers negligible $\Delta\langle L_y\rangle_{x_0,\xi_0}$ anywhere, consistent with the perturbation approaching steady state. The blue curve, just below, plots $|\Delta\langle L_y\rangle_{x_0,\xi_0}|_{max}$, showing that a perturbation transient comparable to the optical pulsewidth ($h_\xi/w_{0\xi} \sim 1$) is most effective in maximizing duration of torque. For the case of an energy-limited perturbation, increasing its duration $h_\xi$ may result in decreasing $\Delta\phi_{p0}$. The red dashed curve, using the constraint $(h_x h_\xi)^{-1} \int dx d\xi\, \Delta\phi_p(x,\xi) = 1$, shows this effect, where in this case the most effective perturbation is the shortest.

## APPENDIX B: DETAILED EXPERIMENTAL SETUP

The detailed experimental setup is shown in Fig. B1.

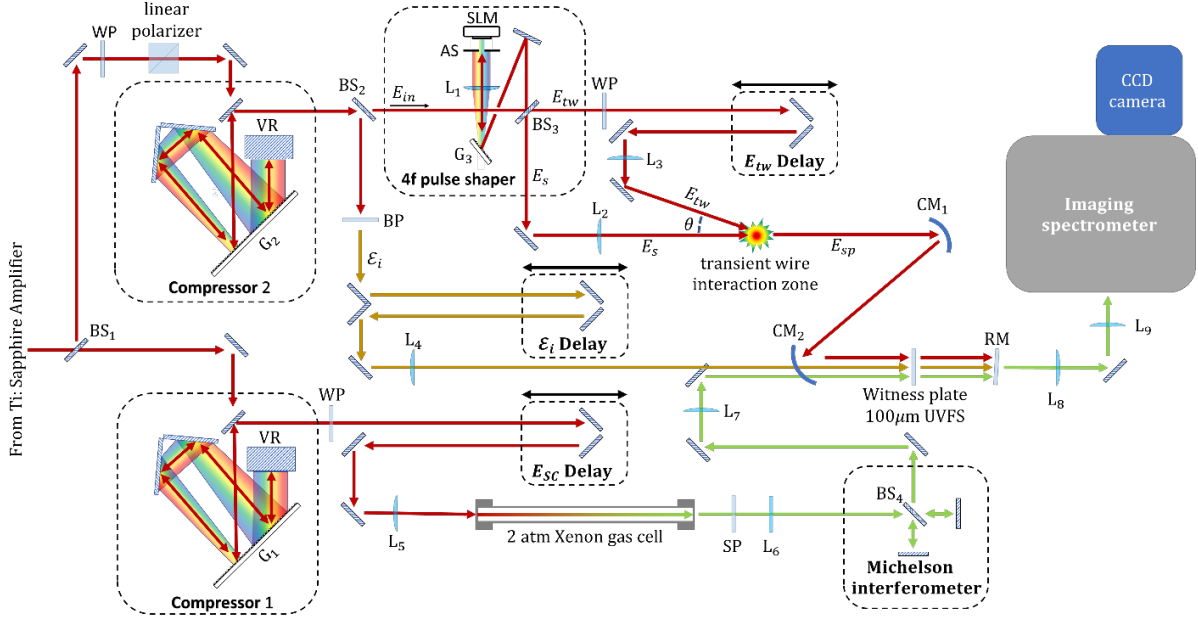

**Figure B1.** Detailed experimental configuration. Compressor 1 adjusts pulse for $4f$ pulse shaper, TG-SSSI spatial interferometry pulse $\mathcal{E}_i$ and transient wire pulse $E_{tw}$. Compressor 2 adjusts pulse for TG-SSSI probe and reference supercontinuum pulses [16]. CM: concave mirror, BP: bandpass filter 3nm FWHM, 800nm center, AS: adjustable slit, SLM: spatial light modulator, WP: λ/2 waveplate, SP: short pass filter, transmits below 750nm, VR: vertical retroreflector, RM: pump/interferometric reference rejection mirror, G: grating, L: lens.

## APPENDIX C: PROPAGATION SIMULATIONS

3D+1 (3 space dimensions plus time) simulations of non-paraxial pulse propagation were performed using our UPPE (unidirectional pulse propagation equation) [41, 42] implementation called YAPPE (yet another pulse propagation effort) [43]. For linear propagation in a dispersive medium in the group velocity frame, YAPPE solves a system of ordinary differential equations,

$$\frac{\partial}{\partial z}\tilde{E}_{k_x,k_y}(\omega, z) = iK_z(\omega, k_x, k_y)\tilde{E}_{k_x,k_y}(\omega, z),  \tag{C1a}$$

$$\tilde{E}_{k_x,k_y}(\omega, z) = \tilde{E}_{k_x,k_y}(\omega, z=0)\exp(iK_z(\omega, k_x, k_y)z). \tag{C1b}$$

Here, $\tilde{E}_{k_x,k_y}(\omega,z) = \mathcal{F}_{x,y,\tau}\{E(x,y,\tau;z)\}$ is the 3D Fourier transform of the spacetime field $E(x,y,\tau;z)$, where $\tau = t - z/v_g(\omega)$ is local time in the pulse frame of reference, $\omega$ is the angular frequency, $v_g(\omega)$ is the frequency-dependent group velocity in the medium, and $K_z(\omega, k_x, k_y) = ((\omega/v_g(\omega))^2 - (k_x^2 + k_y^2))^{1/2} - \omega/v_g(\omega)$, which models diffraction and dispersion. The transverse wavenumber $(k_x, k_y)$ indexes the system of equations (Eq. (C1)), which are numerically solved. To recover the field in the spacetime domain, a 3D inverse Fourier transform is performed, $E(x,y,\tau;z) = \mathcal{F}^{-1}_{k_x,k_y,\omega}\{\tilde{E}_{k_x,k_y}(\omega,z)\}$.

# Spatiotemporal torquing of light: Supplemental material

S.W. Hancock, S. Zahedpour, A. Goffin, and H.M. Milchberg

*Institute for Research in Electronics and Applied Physics, University of Maryland, College Park, Maryland 20742, USA*

## 1. Spatiotemporal phase perturbations localized in time and space

We consider the effect of a localized pure phase perturbation $\Gamma(x,\xi) = \exp(i\Delta\Phi_p(x,\xi))$ centered at $(x_0, \xi_0)$, where

$$\Delta\phi_p(x,\xi) = \Delta\phi_{p0} \exp\left[-\frac{(x-x_0)^2}{h_x^2} - \frac{(\xi-\xi_0)^2}{h_\xi^2}\right], \tag{S1}$$

on both Gaussian and STOV pulses. We first consider the Gaussian at $z=0$,

$$A_G(x,\xi) = A_0 \exp\left(-\frac{x^2}{w_{0x}^2} - \frac{\xi^2}{w_{0\xi}^2}\right), \tag{S2}$$

Applying Eqn. (3) from the main paper,

$$\Delta\langle L_y\rangle = \langle L_y\rangle_{sp} - \langle L_y\rangle_s = iu_{sp}^{-1}\int dx\, d\xi\, \left[|A_s|^2|\Gamma|^2 L_y\Delta\phi_p + |A_s|^2\left(|\Gamma|^2 - \frac{u_{sp}}{u_s}\right)L_y\phi_s\right], \tag{S3}$$

yields a change in transverse OAM per photon

$$\Delta\langle L_y\rangle_G = \frac{8\Delta\phi_{p0}\bar{x}_0\bar{\xi}_0\eta_x\eta_\xi}{\sigma_x^3\sigma_\xi^3}\left(\alpha + \frac{\beta_2}{\alpha}\right)\exp\left(-\frac{2\bar{x}_0^2}{\sigma_x^2} - \frac{2\bar{\xi}_0^2}{\sigma_\xi^2}\right). \tag{S4}$$

Here $\alpha = w_{0\xi}/w_{0x}$, $\eta_x = h_x/w_{0x}$, $\eta_\xi = h_\xi/w_{0\xi}$, $\bar{x}_0 = x_0/w_{0x}$, $\bar{\xi}_0 = \xi_0/w_{0\xi}$, $\sigma_x = \sqrt{1 + 2h_x^2/w_{0x}^2}$ and $\sigma_\xi = \sqrt{1 + 2h_\xi^2/w_{0\xi}^2}$. Similarly, for a transverse OAM-carrying STOV pulse,

$$A_{STOV}(x,\xi) = \left(\frac{\xi}{w_{0\xi}} + i\frac{x}{w_{0x}}\right)A_G(x,\xi), \tag{S5}$$

we find

$$\Delta\langle L_y\rangle_{STOV} = \Delta\phi_{p0}\frac{8\bar{x}_0\bar{\xi}_0\eta_x\eta_\xi}{\sigma_x^5\sigma_\xi^5}\left(\alpha + \frac{\beta_2}{\alpha}\right)\exp\left[-\frac{2\bar{x}_0^2}{\sigma_x^2} - \frac{2\bar{\xi}_0^2}{\sigma_\xi^2}\right] \\ \times \left[\eta_x^2 + \eta_\xi^2 + 8\eta_x^2\eta_\xi^2 - 1 + 2\frac{\sigma_\xi^2}{\sigma_x^2}\bar{x}_0^2 + 2\frac{\sigma_x^2}{\sigma_\xi^2}\bar{\xi}_0^2\right] \tag{S6}$$

Figure S1 plots $\Delta\langle L_y\rangle_G$ and $\Delta\langle L_y\rangle_{STOV}$ vs. $(x_0, \xi_0)$. Both the Gaussian and STOV plots exhibit a 4-lobed structure with respect to spatiotemporal centre of energy.

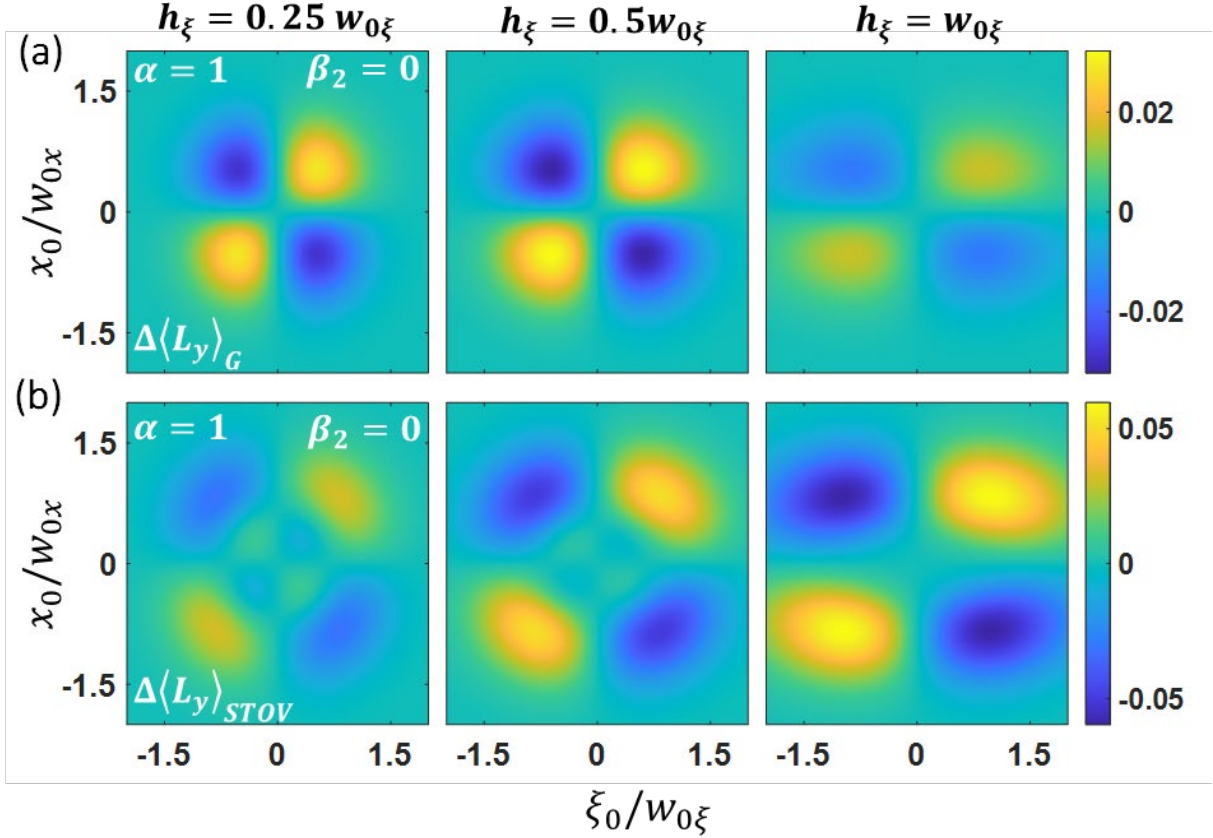

**Figure S1.** (a) $\Delta\langle L_y\rangle_G$ from Eqn. (S4); (b) $\Delta\langle L_y\rangle_{STOV}$ from Eqn. (S6). For the Gaussian phase perturbation $\Gamma(x,\xi) = \exp(i\Delta\phi_p(x,\xi))$, where $\Delta\phi_p(x,\xi) = \Delta\phi_{p0}\exp(-(x-x_0)^2/h_x^2 - (\xi-\xi_0)^2/h_\xi^2)$. Here, $h_x = 0.25w_{0x}$, $\beta_2 = 0$, and $\alpha = w_{0\xi}/w_{0x} = 1$.

## 2. Spatiotemporal amplitude perturbations localized in time and space

Now, we consider the non-energy-conserving amplitude perturbation $\Gamma(x,\xi) = 1 - \Gamma_0\exp(-(x-x_0)^2/h_x^2 - (\xi-\xi_0)^2/h_\xi^2)$ centred at $(x_0,\xi_0)$. Applying this perturbation to a Gaussian pulse, with $\phi_s = \text{const}$, yields $\Delta\langle L_y\rangle = 0$. However, for pulses already carrying a spatiotemporal phase, an analytic result such as Eq. (S6) is difficult. Figure S2(a) plots a numerical calculation of $\Delta\langle L_y\rangle_{STOV}$, as a function of perturbation location $(x_0,\xi_0)$, for the $l=1$ STOV pulse of Eq. (5b), with $w_{0x} = w_{0\xi} = 1$, $h_x = h_\xi = 0.25w_{0x}$, and $\Gamma_0 = 1$. The dashed red circle is the spatiotemporal location of pulse peak intensity. Figure S2(b), shows the change in OAM from a non-transient amplitude perturbation ($h_\xi \to \infty$). Note that for $h_\xi = 0.25w_{0\xi}$, $|\Delta\langle L_y\rangle| = 0.028$, while for the case $h_\xi \to \infty$, $|\Delta\langle L_y\rangle| = 0.12$, as more energy is removed by the perturbation. It is also interesting that the 4-lobed structure from the amplitude perturbation is complementary to that of the Gaussian phase perturbations in Fig. S1. Figure S2(c) examines how the change in OAM varies with $h_x$ for $h_\xi = \infty$, where we see that outside of the contour of peak intensity of the field, $\Delta\langle L_y\rangle$ can be slightly positive while inside the peak intensity contour $\Delta\langle L_y\rangle$ only decreases as $h_x$ increases.

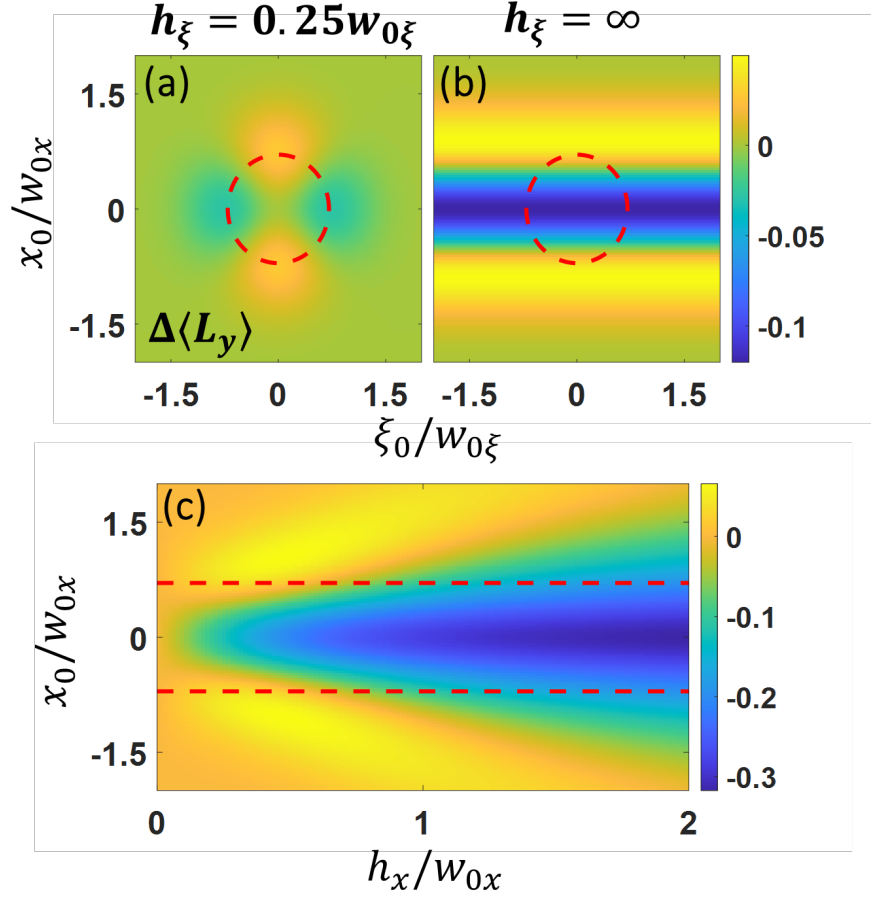

**Figure S2.** (a) Plot of $\Delta\langle L_y\rangle_{\text{STOV}}$ from the amplitude perturbation $\Gamma(x,\xi) = 1 - \Gamma_0\exp(-(x-x_0)^2/h_x^2 - (\xi-\xi_0)^2/h_\xi^2)$, with $w_{0x} = w_{0\xi} = 1$, $h_x = h_\xi = 0.25w_{0x}$, and $\Gamma_0 = 1$. (b) Same as (a) except for $h_\xi \to \infty$. (c) Plot of $\Delta\langle L_y\rangle_{\text{STOV}}$ as a function of $h_x$ for $w_{0x} = w_{0\xi} = 1$, $\Gamma_0 = 1$, and $h_\xi \to \infty$. The dashed red lines denote the contour of peak intensity of $A_{\text{STOV}}(x,\xi)$.

3